\documentclass[11pt,a4paper]{article}

\usepackage[a4paper,margin=1.5in]{geometry}
\usepackage{xltabular}
\usepackage[T1]{fontenc}
\usepackage[utf8]{inputenc}
\usepackage{tocbibind}
\usepackage{url}
\usepackage{apacite}
\usepackage{graphicx}
\usepackage{comment}
\usepackage{float}
\usepackage[font=small,labelfont=bf,labelsep=period]{caption}
\usepackage{makecell}
\usepackage{mathtools}
\usepackage{commath}
\usepackage{booktabs}
\usepackage{multirow}
\usepackage{adjustbox}
\usepackage{enumitem}
\usepackage{bbold}
\usepackage{xcolor}
\usepackage{amssymb}
\usepackage{amsmath}
\usepackage{mathtools}
\usepackage{dutchcal}
\usepackage{tabularx}
\usepackage{pdflscape}
\usepackage{subcaption}
\usepackage[bottom]{footmisc}
\usepackage{helvet} 
\fontfamily{phv}\selectfont

\usepackage{rotating}

\usepackage{color}
\usepackage{soul}
\sethlcolor{yellow}

\usepackage{placeins}
\let\Oldsection\section
\renewcommand{\section}{\FloatBarrier\Oldsection}

\RequirePackage{fix-cm} 
\DeclareMathSizes{12}{12}{10}{10}
\usepackage[doublespacing]{setspace}

\newcommand{\begintext}{%
        \setcounter{page}{1}
     }

\usepackage[ruled, vlined, linesnumbered]{algorithm2e}
\usepackage{etoolbox}
\AtBeginEnvironment{algorithm}{\setstretch{1.1}}
\SetCommentSty{\footnotesize\texttt}

\newboolean{showcomments}
\setboolean{showcomments}{true}
\ifthenelse{\boolean{showcomments}}
{ \newcommand{\mynote}[3]{
    \fbox{\bfseries\sffamily\scriptsize#1}
    {\small$\blacktriangleright$\textsf{\textit{\color{#3}{#2}}}$\blacktriangleleft$}}}
{ \newcommand{\mynote}[3]{}}

\newcommand{\beginappa}{%
        \setcounter{page}{1}
        \setcounter{table}{0}
        \renewcommand{\thetable}{A.\arabic{table}}%
        \setcounter{figure}{0}
        \renewcommand{\thefigure}{A.\arabic{figure}}
        \setcounter{subsection}{0}
        \renewcommand{\thesubsection}{\mbox{A.\arabic{subsection}}}
     }

\title{Tournaments, Contestant Heterogeneity and Performance\footnote{We are grateful to Jean-Michel Benkert, Riccardo di Francesco, Moritz Janas, Tommy Krieger, Alex Krumer, Michael Lechner, Igor Letina, Maurizio Strazzeri, and the participants of the Quantitative Economics Workshop 2023 at the University of St.Gallen, the REUS workshop 2023 and the SES 2024 for their useful comments and suggestions.}}
\author{Enzo Brox\footnote{University of St.Gallen: enzo.brox@unisg.ch} \and Daniel Goller\footnote{University of Bern: daniel.goller@unibe.ch}}

\date{\today}
\begin{document}

\maketitle
\thispagestyle{empty}

\begin{abstract}
 Tournaments are frequently used incentive mechanisms to enhance performance. In this paper, we use field data and show that skill disparities among contestants asymmetrically affect the performance of contestants. Skill disparities have detrimental effects on the performance of the lower-ability contestant but positive effects on the performance of the higher-ability contestant. We discuss the potential of different behavioral approaches to explain our findings and discuss the implications of our results for the optimal design of contests. Beyond that, our study reveals two important empirical results: (a) affirmative action-type policies may help to mitigate the adverse effects on lower-ability contestants, and (b) the skill level of potential future contestants in subsequent tournament stages can detrimentally influence the performance of higher-ability contestants but does not affect the lower-ability contestant.
 
\end{abstract}

\textbf{JEL-Codes:} J33, M52, D02, J78\par
\par
\newpage

\section{Introduction}
\begintext

Tournaments are widely used incentive mechanisms in various domains, ranging from science and innovation to business, politics, and sports \cite{Baye.1993, Terwiesch.2008, Gurtler.2015}. They typically involve two or more contestants competing for a reward, with the winner determined by their relative performance \cite{Corchon.2018, Lazear.2018}. Tournaments also have in common that they are intentionally designed by a principal who seeks to optimally select parameters such as contest rules, prize structures, and contestant compositions to align with her specific objectives \cite{Che.2003, Letina.2023}.
To guide contest designers towards the optimal design of a contest, this paper investigates how an essential parameter in this design -- the relative abilities of the contestants -- affects performance \cite{Lazear.1981}.\par

Economic theory predicts that heterogeneity in ability among contestants is detrimental to overall performance \cite{Rosen.1986}. In a two-player contest, the argument typically unfolds as follows: as skill-level disparities increase, the contestant with lower ability has reduced incentives to exert high effort due to a diminished probability of winning the reward, leading to lower performance levels. 
The contestants with higher ability, in turn, reduce effort because they anticipate the drop in performance of the contestants with lower ability.
While there are a few empirical studies that show the individual response of the lower-ability contestant (e.g., \citeA{Brown.2011}), there is, to our knowledge, no field study that can credibly demonstrate the individual responses of both higher-ability and lower-ability contestants. Such an analysis, however, holds significant importance in accounting for the diverse objectives of contest designers.\par

Consider two principals, A and B. Principal A is planning to conduct a public tender or an innovation tournament, while Principal B manages a sales team. The objective functions of Principal A and B differ. While Principal A seeks the best performance among all contestants, Principal B focuses on overall performance or aims to boost those not yet performing at the highest level. By considering how the relatively more skilled and the relatively less skilled contestants respond to heterogeneity in ability, our study guides contest designers with different objective functions toward the optimal contest design. We also address two additional relevant questions for contest designers related to contestant heterogeneity in ability: First, we assess the impact of providing a head start to the lower-ability contestant, as evident in many affirmative action policies, and its impact on the performance of heterogeneously skilled contestants. Second, we consider the multi-stage nature of a tournament-like situation, as frequently seen in public tenders or innovation competitions, and investigate whether the strength of potential future competitors affects performance in the current stage.\par

To investigate how heterogeneity in ability among contestants affects individual performance, we use field data from professional darts tournaments.
Darts tournaments provide an ideal setting for this purpose as they provide accurate information about contestants' abilities, and the tournament design allows us to quantify ability heterogeneity clearly. Furthermore, the context is particularly attractive because individual performance is precisely quantifiable, and the contestant's performance is not (directly) affected by the other contestant. Notably, the performance is observed in a situation that the contestants are used to, with high stakes and few influencing factors.\par

For our empirical analysis, we rely on the design of darts tournaments as a source of quasi-random variation in ability heterogeneity among contestants. This variation arises from the draw of contest pairs, in which lower-ability contestants are randomly paired with higher-ability contestants. We provide results using standard linear regression frameworks and use causal machine learning methods to investigate non-linearities in the effect of ability heterogeneity on the individual performance of contestants.
To complete the picture, we show how the individual responses to ability heterogeneity affect the final result and other contest outcomes. We run a battery of robustness checks to confirm the econometric identification and to test the robustness of our findings in the face of variations in the measurement of key variables.\par

The findings can be summarized as follows. In line with the theoretical predictions, as contestant ability heterogeneity increases, the performance of the lower-ability contestant decreases. While previous studies (e.g., \citeA{Brown.2011}) suggest that the negative effect on the performance of the lower-ability contestant is only observed when the heterogeneity among contestants is large enough, we observe a negative effect already at moderate ability differences. Thus, our results support the conventional wisdom of a "discouragement effect" for lower-ability contestants arising from heterogeneity in contestant ability \cite{Drugov.2022}.\par  

Contrary to the theoretical predictions, we do not observe a negative performance effect for the higher-ability contestant. As the heterogeneity of contestants' abilities increases, the higher-ability contestant does not decrease performance. Instead, we observe a moderate increase in performance. Remarkably, in the first half of the contest, when ex-ante ability heterogeneity closely determines the winning probabilities, we find a substantial positive effect. 
We also show that the positive effect on the performance of the higher-ability contestant is driven by the contestants with the highest abilities in our sample. Overall, the individual responses to increasing ability heterogeneity in the aggregate lead to a higher likelihood that the higher-ability contestant wins the contest -- beyond what the pure ability differences would suggest.\par

Our findings align with several recent theoretical articles on optimal contest design, which argue that the highest degree of competitiveness is not necessarily the optimal choice for a contest designer \cite{Fang.2020, kirkegaard.2023, Letina.2023}. We provide a theory-guided discussion about potential mechanisms that can explain our findings. We show that out-of-equilibrium beliefs and a differential evaluation of rewards are unlikely to explain our results. Two potential explanations for the decline in the performance of higher-ability contestants with decreasing heterogeneity among contestants are choking under pressure \cite{Baumeister.1984, Ariely.2009, Harb.2019} and anticipated regret \cite{Hyndman.2012}. We use a simple theoretical model to show that incorporating choking under the competitive pressure ("of close contests") of the higher-ability contestant or anticipating regret from losing against a much weaker opponent can account for the observed pattern. \par 

Having documented the negative effect of heterogeneity in ability among contestants for the lower-ability contestant, we extend our analysis to investigate whether an affirmative action type policy that reduces the effective degree of heterogeneity among contestants can counteract the observed effect \cite{Chowdhury.2023}. We exploit a unique feature of darts competitions, i.e., a built-in advantage and conditional random variation in this 'head start' advantage to show that a head start for the lower-ability contestant does have the proposed incentive effect for the lower-ability contestant without adverse effects on the individual performance of the higher-ability contestant. We also find that the effect of a head start is stronger for larger levels of ex-ante heterogeneity among contestants.\par

Finally, we investigate the impact of potential future competitors on the current performance. In many settings, tournaments are multi-stage events where contestants compete for immediate rewards and the chance to earn additional rewards in later stages \cite{Sheremeta2010, Segev2014}. This bears the possibility that the strength of (potential) competitors in future stages of the tournament has shadow effects on the performance in the current stage. To analyze this, we use an instrumental variable approach that exploits unexpected upsets in combination with the schedule of contests. We find that the higher-ability contestants exhibit forward-looking behavior: With the increasing strength of the future opponent, the higher-ability contestants significantly reduce their performance in the current contest, with meaningful implications for the likelihood of winning the current contest. The lower-ability contestant does not show a performance response. \par

Our results have several practical implications regarding the optimal design of contests. Consider Principal A and Principal B from the example above again. Principal A wants to conduct a public tender, while Principal B wants to motivate her sales team. First, our results show that the optimal degree of heterogeneity depends on the principal's objective function. Low ability heterogeneity may benefit lower-ability contestants' performance, but a higher level of heterogeneity is not detrimental to the performance of higher-ability contestants. In particular, it is (most) performance-enhancing for the contestants who are expected to perform the best and are, therefore, very important to Principal A. Second, for Principal B, biasing contests by giving the lower-ability contestant a head start could be an interesting way to mitigate adverse incentive effects arising from heterogeneity in ability among contestants since it does not harm the performance of the higher-ability contestant but enhances the performance of the lower-ability contestant. Third, in multi-stage tournaments, such as public tenders or innovation competitions, it may be necessary, especially for Principal A, to consider how to mitigate shadow effects between tournament stages.\par

\subsection{Relation to Literature}

Due to the frequent use of tournaments as incentive mechanisms, a large literature has emerged that investigates how different tournament features affect performance and how to optimally design a tournament (see, e.g., \citeA{Corchon.2018} and \citeA{Lazear.2018}, for surveys). 
With this article, we complement three different strands of literature on elimination tournaments.
First, we contribute to the literature exploring the effects of heterogeneity in contestants' abilities on behavior in tournaments (e.g., \citeA{Konrad.2009a}, for a survey). Second, we contribute to the literature on biased contests (e.g., \citeA{kirkegaard.2012, Drugov.2017}), and third to the literature on multi-stage tournaments (e.g., \citeA{Sheremeta2010, Fu.2012}). Below, we summarize the relevant studies and clarify our contributions with regard to the literature. \par

Differences in relative abilities among contestants are typically considered discouraging effort provision and performance.\footnote{Note here that this result is not universal. There are instances in which greater heterogeneity implies higher total effort. See, for example, \citeA{Drugov.2017, Drugov.2022} and \citeA{Gurtler.2015}.} 
Experimental studies support the argument of a discouragement effect in rank-order tournaments \cite{Bull.1987, Schotter.1992}.\footnote{For an overview on experimental evidence of a discouragement effect in different contest-type situations, see \citeA{Dechenaux.2015}.} Only a few studies have used field data due to a lack of information on ability and performance and a suitable identification strategy that allows for a causal interpretation of the obtained results.\footnote{\citeA{Boudreau.2011} study the effects of the number of competitors on effort provision on online platforms, but does not take ability differences between contestants into account. \citeA{Boudreau.2016} investigate heterogeneity in the response to an increase in the number of competitors with respect to the contestant's ability. \citeA{Ammann.2016} show that CEOs who experienced a status shock exhibit an encouraging effect on competitors' performance.} \citeA{Chan.2014}, using data from sales departments, find that heterogeneity in worker ability improves firm performance under team-based compensation and hurts under individual-based compensation. \citeA{Gross.2020} provides evidence for a non-linear effect of the intensity of competitions on creative production using data from commercial logo design competitions. While some degree of competition, created by interim ratings, is conducive to creativity, very high-intensity levels reduce creative performance. Closest to our study, \citeA{Brown.2011} exploits the (unexpected) absence of a superstar in elite golf tournaments and shows that the presence of superstars can hurt the performance of non-superstar competitors. The related literature on dynamic contests argues that similar incentive effects can arise due to ahead-behind asymmetries in contests with symmetric contestants \cite{Malueg.2010, Gauriot.2019}.\par

We complement previous studies in several ways. First, we use field data from professional darts tournaments and examine the individual responses of higher- and lower-ability contestants to contestant heterogeneity.\footnote{Several studies use data from the professional sports industry to test whether the behavior of individuals violates standard economic assumptions. E.g., \citeA{Pope.2011} use data from professional golf tournaments to test prospect theory and \citeA{Genakos.2012} use data from weightlifting competitions to investigate risk-taking behavior in tournaments. For studies using data from darts competitions, see, for example, \citeA{Tibshirani.2011}, \citeA{Klein.2020}, \citeA{Goller.2023}, and \citeA{Scholten.2023}.} This is important to draw a more nuanced picture of contestant ability heterogeneity's effect and allow speaking to tournament designers' different potential target functions. Our results question the general perception that heterogeneity among contestants is necessarily detrimental to performance. Second, in contrast to \citeA{Brown.2011}, we use a continuous measure for contestant heterogeneity that allows us to evaluate the effect of different doses of skill heterogeneity. Another advantage of this approach is that we do not have to rely on a particular and potentially subjective definition of a superstar but derive information on past performance to measure ability heterogeneity objectively. Third, we discuss potential reasons for the observed deviation from the predicted behavior for higher-ability contestants. We show that a simple model that accounts for choking under competitive pressure of higher-ability contestants or the anticipation of regret from losing against a much weaker opponent is in line with our findings.\par 

Our study also contributes to the literature on biased contests. Some theoretical studies have investigated the incentive effects of biased contests, in which a subset of the contestants is given a "head start" \cite{kirkegaard.2012, Drugov.2017}. Although policies that bias contests are widely used (e.g., to fight discrimination), empirical evidence on how they affect performance is limited due to the difficulty of measuring performance \cite{Calsamiglia.2013, Leibbrandt.2018}. In a recent study, \citeA{Saguer.2023} use an auction design to show that favoring the lower-ability contestant diminishes the discouragement effect for the lower-ability contestant. However, the higher-ability contestant's less aggressive bidding behavior offsets the overall benefits. \citeA{Franke.2012} studies handicap rules in amateur golf tournaments and finds that contestants perform better, underscoring rules that account for differences in ability. We contribute to this literature by providing causal evidence using field data on the effect of head starts for lower-ability contestants in rank-order tournaments on the individual performance of higher-ability and lower-ability contestants. We also evaluate their relation to contestants' ability heterogeneity.\par

Lastly, our study relates to the literature on incentive effects in multi-stage tournaments. \citeA{Brown.2014} present a theoretical model that shows how the presence of high-ability contestants in future stages of a tournament can reduce the effort of both contestants. They also show that an increasing strength of future contestants results in a lower probability of winning for the higher-ability contestant in the current stage since the higher-ability contestant reduces the effort to a larger degree than the lower-ability contestant. They use match outcomes from professional men's tennis tournaments to show that the higher-ability contestant is less likely to win a contest with the increasing strength of competitors in future stages. \citeA{Kleinknecht.2021} replicate the result for women and find small gender differences. \citeA{Lackner.2020} use data from professional and semi-professional basketball tournaments and information on personal fouls to measure effort provision. They find that also in a teamwork setting, contestants exhibit forward-looking behavior, since with an increasing strength of future competitors, effort provision decreases in the current stage.\par 

We contribute to this literature by estimating the individual responses to increasing contestants' strength in future stages of a tournament and thus by investigating the underlying mechanisms of the model in \citeA{Brown.2014}. In line with previous studies, the higher-ability contestant is less likely to win the contest with the increasing strength of competitors in future stages. When looking at the individual performance of the contestants, we find results that, in relative terms, align with the theoretical model. In absolute terms, we do not find evidence for a forward-looking behavior of the lower-ability contestant. The result is driven by a performance decrease of the higher-ability contestant.\par

\section{Setting}\label{sec:setting}

We use data from professional darts tournaments to empirically test how contestant ability heterogeneity affects the individual contestants' performance. Darts has received a large increase in public and media interest over the past decade and provides a perfect environment to answer our research questions for several reasons. First and most importantly, we can observe individual performance in a contest without direct interaction between contestants. Second, we have precise information about the contestants' abilities, and due to the single-elimination tournament design, heterogeneity in the ability can be easily determined. Third, we can monitor individuals who perform their daily work in high-stakes environments and are familiar with contests. Fourth, darts contests take place in a standardized, laboratory-like environment and are played according to clear, non-subjective rules. \par

Darts tournaments are primarily played in single-contest knockout tournaments and are held according to the rules of the Darts Regulation Authority. Depending on the tournament, 16 to 96 participants compete for the tournament victory in four to seven stages. In the first stage of each tournament, participants are randomly assigned to a contest, while the tournament's structure determines future fixtures. This draw is made with the restriction that a predetermined number of participants with higher rankings according to a world ranking list (called "order-of-merit") are anchored in the schedule and meet (randomly determined) opponents with lower rankings. In addition, some tournaments have preliminary stages where the top-ranked participants receive a bye stage -- in this case, they are assigned to the second stage. All contests are staged one after the other. The incentives in the tournament are determined by a multi-prize scheme, i.e., the winner gets most of the total prize money and everybody else an amount that is lower the earlier they are eliminated from the tournament. \par 

In each contest, two contestants meet to determine which contestant advances to the next stage while the other is eliminated from the tournament. Contests are played in a best-of-\textit{k} legs format, meaning that to advance to the next stage, a contestant must win $\dfrac{k+1}{2}$ legs.\footnote{Some tournaments use a best-of-\textit{m} sets format, where each set consists of five legs. Odd numbers are used for \textit{k} (and \textit{m}).}
Whoever collects exactly 501 points before his opponent and scores a "double" (a particular area on the edge of the dartboard that doubles the points from the throw) with the last throw has won a leg -- the trailing opponent cannot catch up. Throughout the contest, the two contestants make their moves sequentially, with three darts thrown on each move. The maximum score in each move is 180 points, i.e., three darts, each of which scores the maximum of 60 points. \par 

\section{Data}\label{sec:data}

The data set contains 4776 contests, including over 400 players in about 150 darts tournaments from 2010 to 2020. The data set was generated from the software \textit{dartsforwindows}.\footnote{\textit{dartsforwindows} is a computer scoring software for darts tournaments that is used for live scoring in darts tournaments. While it is used in most professional tournaments to record live scoring and provide live, up-to-date statistics, detailed log files of all games are stored for documentation and made available online. The online database is supervised by a former professional darts player and categorized into amateur and professional tournaments, whereby we only use professional tournaments due to the lack of adequate measures of ability in amateur tournaments. We have excluded (a) tournaments for which we do not have the full schedule, (b) tournaments in which contests are not played on a single board and therefore do not take place strictly sequentially, (c) contests featuring contestants for which we cannot observe past performance. Moreover, we stopped collecting data at the beginning of the COVID-19 pandemic since tournaments were canceled, rescheduled, or conducted with fewer or no spectators and potentially underlying different incentives.} To better distinguish contestants by ability in each contest, we label the relatively more skilled contestant (higher-ability contestant) as the favorite and the relatively less skilled contestant (lower-ability contestant) as the underdog and use those descriptions interchangeably. The data set includes contest outcomes, such as the winner of a contest and the performance of the contestants in the contest. It also includes information on who starts in the first leg, when the contest is scheduled, and the contestants. For an overview of the essential variables for our main analysis, see Table \ref{tab:table_main_desc}; Appendix Table \ref{tab:table_app_desc} contains the complete set of variables.\par

\textbf{Outcome variables} For any contest, we observe the score of each contestant for each move in each leg.  
We take this information to measure individual performance by aggregating over the first three moves of a contestant in a leg. This 3-darts average for the first nine darts (hereafter called "performance") is a frequently used performance metric in professional darts and offers some significant merits for answering our research questions. More specifically, it allows us to quantify the absolute, non-subjective performance of the individual without interference from the opponent or the environment. Within the first nine darts, the technical requirement to finish a leg with a double is essentially irrelevant, and there are few if any opportunities for strategic decisions that could blur the absolute performance \cite{Klein.2020}.\footnote{The prevailing strategy within the first nine darts is to score the maximum number of points possible. Strategic choices solely arise as soon as a contestant 'sets' the remaining number of points to be scored to a certain value by a throw that is not aimed at the maximum point yield. Note that a contestant needs at least nine darts to complete a leg, while legs played with nine darts are extremely rare. In our sample, the median number of darts needed to finish a leg is 15. However, we conduct robustness checks using only the 3-darts average for the first six darts as an alternative outcome variable.}\par

\begin{figure}[H]
      \caption{Outcome and Treatment Variables}
      
    \begin{subfigure}{.5\textwidth}
      \centering
      \includegraphics[width=.97\textwidth]{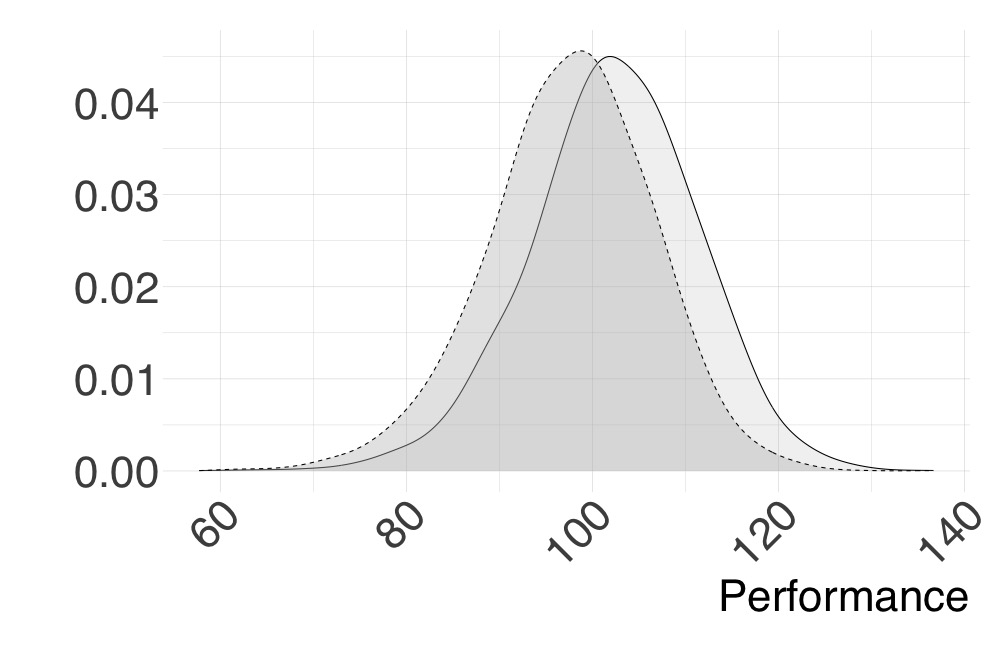}
      \caption{Performance higher- and lower-ability}
      \label{fig:desc_out}
    \end{subfigure}
    \hfill
        \begin{subfigure}{.5\textwidth}
      \centering
      \includegraphics[width=.99\textwidth]{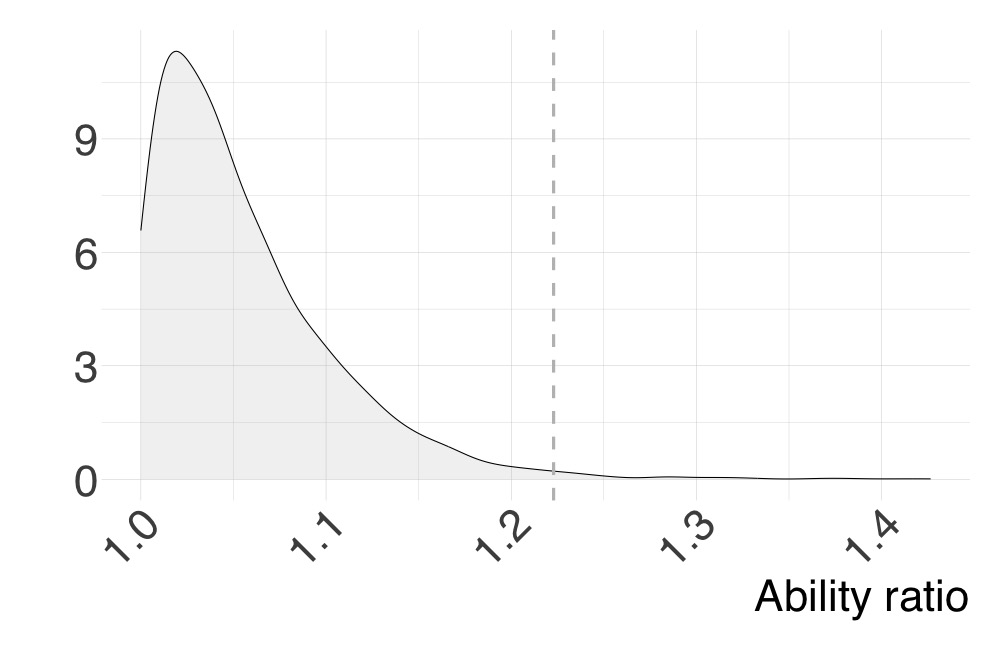}
      \caption{Treatment variable distribution}
      \label{fig:treat_desc}
    \end{subfigure}
    \center{
	\footnotesize Notes: a) Performance distributions for the lower-ability (dark) \& the higher-ability (light) contestants.}   \\
	\footnotesize ~~~~~~~~ Performance is the 3-darts average (first nine darts). b) Distribution of the treatment variable, ~ \\
	\footnotesize ~~~~~~~~~~~ defined as the higher-ability contestant ability divided by the lower-ability contestant ability. The \\
 	\footnotesize ~~~~~~~~ broken line marks the 99\% quantile. Distributions are smoothed using a local constant kernel. ~~ \\
\end{figure}

In Figure \ref{fig:desc_out}, we show the distribution of our primary outcome variables. The solid line shows the density of the realized performance of the higher-ability contestants, and the broken line depicts the lower-ability contestants' performance. While both outcome variables are distributed from about 60 to 135, it is already evident from Figure \ref{fig:desc_out} that the lower-ability contestants also realize a lower performance. This is also documented in Table \ref{tab:table_main_desc} with mean values of 102.254 for the higher-ability and 97.595 for the lower-ability contestants. \par 

To complement our results, we use the average performance of both contestants in a contest, with a mean value of 99.924, as well as a binary indicator of whether the higher-ability contestant won a contest, which is the case in 66.5 percent of all contests. Supplementary outcome variables are the length of the contest, which amounts on average to about 79 percent of the maximum possible number of legs (\textit{k}) in the respective stage, and the \textit{number of maximum performance}, averaging at about six times, per contest. This corresponds to the number of times a contestant scores 180 points in one move.\par

\begin{table}[H]
		\centering
		\caption{Descriptive statistics} 
		\label{tab:table_main_desc}
		\begin{tabular}{l r r r} 
			\toprule \toprule

            \multicolumn{4}{l}{\textit{Panel A: Outcomes}} \vspace{0.7em}        \\ 
			Favorite wins                                & ~~~ &  0.665 &           \\
		    Favorite 3-darts average (performance)       & &  102.254    & (9.057)   \\
		    Underdog 3-darts average (performance)       & &  97.595     & (8.879)   \\
		    Mean 3-darts average (performance)           & &  99.924     & (6.859)   \\ 
            Contest length (proportion of legs needed)   & &  0.790      & (0.159)   \\
            Number of max. performance (Score = 180) in contest & & 5.921 & (4.186)  \\ 
            \vspace{-0.5em}  \\
            \multicolumn{4}{l}{\textit{Panel B: Treatments}} \vspace{0.7em} \\ 
			Ability ratio (favorite to underdog)         & &   1.055     & (0.049)   \\
            Expected ability (of opponent in next stage) & &  94.695     & (3.963)   \\
   			Underdog starts contest                        & &  0.481      &           \\
            \vspace{-0.5em}  \\
            \multicolumn{4}{l}{\textit{Panel C: Covariates}} \vspace{0.7em} \\ 
            Opponent is known                            & & 0.588       &           \\
   			Favorite starts contest                        & &  0.519      &           \\
      		Favorite world ranking                       & &   0.017     & (0.081) \\
         	Underdog world ranking                       & &    0.046    & (0.129) \\
            Favorite years playing darts (experience)    & &  21.333     & (8.921)   \\
            Underdog years playing darts (experience)    & &  19.766     & (10.493)  \\
            Favorite performs at home event              & &   0.034     &           \\
            Underdog performs at home event              & &   0.041     &           \\
			\bottomrule \bottomrule
			\multicolumn{4}{l}{\footnotesize Notes: Mean and standard deviations (for non-binary variables). Full descriptive}\\
			\multicolumn{4}{l}{\footnotesize ~~~~~~~~~    statistics can be found in Appendix Table \ref{tab:table_app_desc}.}
		\end{tabular}
	\end{table}

\textbf{Treatment variables} Our main treatment variable, \textit{ability ratio}, is calculated as the ratio of the higher-ability to the lower-ability contestants' ability. Figure \ref{fig:treat_desc} shows the distribution of our main treatment variable. In our main specifications, we use the contestant's 3-darts average over the past two years to calculate the \textit{ability ratio}. Table \ref{tab:table_main_desc} and Figure \ref{fig:treat_desc} indicate substantial heterogeneity in the ability difference. To illustrate this ability heterogeneity: The average higher-ability contestant is about 28 percentage points more likely to win (according to bookmakers' odds) than the average lower-ability contestant.\par

For the extensions of the analysis, two additional treatment variables are used. \textit{Underdog starts contest} is used to investigate whether a head start affects performance and is a binary variable indicating whether the lower-ability contestant starts the contest. \textit{Expected ability}, with a mean value of 94.695 (see Table \ref{tab:table_main_desc}), is the expected ability of the opponent that the contestants expect to meet in the next stage. This is the ability of the higher-ability contestant in the contest that is scheduled to determine the next opponent (the parallel or twin contest). In case this parallel contest is already finished and, consequently, the opponent is already known before the contest, the expected ability is the (realized) ability of the (then known) opponent in the next stage.\par

\textbf{Control variables} We collect additional information on the tournament characteristics, such as the venue, the total purse for the tournament as well as the proportion of prize money that is at stake in the current round and in the remaining future rounds\footnote{The total purse for a tournament shows the importance of the tournament. The distributed money influences incentives twofold: First, winning more money is better than winning less money, and second, for ranking-relevant tournaments, the individual prize money won counts toward the world ranking relevant for qualification to other tournaments. We standardize the total amount of prize money allocated in a tournament $m$ for each year $t$ as $ Prize~money^{standardized}_{mt} = \frac{Prize~money_{mt}-min_t(Prize~money_t)}{max_t(Prize~money_t) - min_t(Prize~money_t)}$, because tournament purses were raised continuously during the observation period. The proportion of prize money at stake in the current and remaining future rounds is calculated as the proportion of prize money distributed for winning the current (winner-takes-all) contest, respectively the proportion of prize money distributed in the remaining future rounds divided by the total tournament purse.}, the draw, the exact order of play, and on the contestants, such as age, hometown, nationality, experience playing darts, and ability measures. The ability measures contain past performance, i.e., 3-darts average over the past two years (hereafter "ability") and the world ranking position (hereafter "ranking" or "world ranking"). The world ranking is recorded at the end of the previous year and is determined by the prize money collected over the past two years. We re-scale this ranking between 0 and 1 (from best to worst), with mean values of 0.017 for the higher-ability and 0.046 for the lower-ability contestants. Experience is measured by the number of years ago the participant started playing darts. The contest information is used to construct additional variables, such as whether the opponent in the next stage is already known at the time of the contest, which is the case in 58.8 percent of all contests, or whether the contestants compete in a "home event," which for the 3.4 (4.1) percent of contestants with higher- (lower-) ability is defined as having a hometown within 100 km of where the tournament is held.\par 

\section{Empirical Strategy}\label{sec:emp_strat}

\subsection{Contestant Heterogeneity}
We analyze the effect of increasing contestant heterogeneity on the contestants' performance. We rely on idiosyncratic variations in the ability ratio of two contestants assigned to a contest stage. The identifying assumption requires no confounding factors influencing both the treatment, i.e., the ability ratio of the two contestants, and the outcome, i.e., performance in the contest. The main approach can be formalized to the baseline (linear) model:

\begin{equation*}
    Y_{cst}^{F/U} = \alpha + \beta D_{cst} + \gamma X_{cst} + \epsilon_{cst},
\end{equation*}

\noindent
where $Y_{cst}^{F/U}$ is the performance of the higher-ability contestant (\textit{F}) or the lower-ability contestant (\textit{U}) in contest \textit{c} and stage \textit{s} in tournament \textit{t}. The ability ratio $D_{cst}$ is defined as the higher-ability contestant ability divided by the lower-ability contestant ability. Potential confounding factors that (might) influence $D$ and $Y$ are captured in $X_{cst}$. $\epsilon_{cst}$ might include unobserved confounding influences associated with the contest and the tournament, i.e., $\epsilon_{cst} = \mu_{cs} + \xi_{ct} + e_{cst}$, where $\mu_{cs}$ and $\xi_{ct}$ are stage and tournament fixed effects, such that $e_{cst}$ is an idiosyncratic error term.\par

From the point of view of both the higher-ability contestant and the lower-ability contestant in a given stage, the opponent is randomly assigned. Consequently, we include individual fixed effects for the higher-ability contestant or lower-ability contestant. To capture the individual ability more flexibly, we control for individual factors that might influence ability ratio and performance, such as both the lower-ability and higher-ability contestant's world ranking and the experience of the contestants.\footnote{In the robustness tests, we check alternative specifications for ability, such as to control directly for the ability or using individual-by-year fixed effects.}\par

Moreover, we include "stage fixed effects" because incentives differ across stages of a tournament, which could be correlated with performance, and, at the same time, the ability ratio becomes closer in subsequent stages of a knock-out tournament. The different relevance of the various tournaments, as well as the differential ways to qualify for the tournaments, is taken into account by tournament-by-year fixed effects. In addition, we consider whether participants compete at a venue near their hometown to account for different incentives and the possibility that some tournaments might allow locals to qualify for the tournament through local qualifiers or by receiving a wildcard.\par

The baseline estimations are conducted using a linear regression framework, and standard errors are clustered at the individual level. To investigate the potential non-linearity of the effects, we extend our analysis and apply a method from the causal machine learning literature. Using the non-parametric kernel method for continuous treatment \cite{Kennedy.2017} has the advantage of resolving both the linear additivity and the constant treatment assumption, which are implicitly assumed in a linear regression framework. This is particularly valuable in this case, as it allows us to examine whether the effect is the same for larger and smaller ability differences.\par 

In detail, the method is composed of two steps. In the first step, a pseudo-outcome is constructed:
\begin{equation*}
        \xi(\pi,\mu) = \frac{Y-\mu(X,A)}{\pi(A|X)} \int \pi(A|x)dP(x) + \int \mu(x,A)dP(x),
\end{equation*}

\noindent
where $\pi(A|X)$ and $\mu(X,A)$, the nuisance functions, are estimated by a random forest \cite{Breiman.2001}, a very flexible and well-performing nonparametric estimator. The pseudo-outcome $\xi(\pi,\mu)$ is free from confounding influences and doubly robust, meaning that (at least) one of the two nuisances needs to be consistent, not both. 
The average potential outcome for the respective treatment levels is estimated in a second step by a non-parametric kernel regression of the pseudo-outcome on the (continuous) treatment variable as $E(Y^a)=E(\xi(\pi,\mu|A=a))$. The bandwidths of the kernels are determined in a data-driven approach using a cross-validation method.\par

\subsection{Head start}\label{subsec:ident_hs}

For our analysis of the effect of affirmative action policies, we exploit that the contestant who starts, by the contest's design, has a head start, i.e., a higher probability of winning the contest than the opponent. When starting a leg in darts, you have a built-in advantage because you have more or equal, but never fewer, darts than your opponent to win a leg and the contest \cite{Goller.2023}.\par

Starting in the first leg in darts is not allocated randomly but awarded to the one who wins a one-dart shootout before the contest. The task in the one-dart shootout is to throw one dart as close as possible to the center of the dartboard. Our empirical strategy builds on a conditional independence assumption, that is, controlling for factors that potentially influence both, starting in the first leg and winning the darts match. First, more able darts players are more likely to win the shootout. Therefore, we control for the ability ratio between the two contestants and the contestant's world ranking. Second, mental issues might also play a role, which is why we consider the individual's years of experience playing darts as a proxy for their familiarity with the situation. Third, individual players might handle the situation differently: Some don't care about the head start, others may think they are better off losing the shootout, and others are just very good or bad at the one-dart shootout. To account for this, we include individual fixed effects. Moreover, we recognize that different incentives may arise in tournaments of different relevance or at different stages. We address this by incorporating tournament-by-year and stage fixed effects. We argue that the conditional independence assumption is likely to be fulfilled with this set of conditioning variables. \par

\subsection{Shadow effects}

To investigate potential shadow effects across stages, we focus on the expected ability of the competitor in a subsequent stage. We estimate the effect of increasing strength of potential future opponents on the performance in the current stage with two different empirical strategies. The expected ability of the potential future opponent is defined as the (known) ability of the opponent in the next stage if the next opponent is known. Otherwise, it is the ability of the favorite of the contest that will determine the next opponent.
First, we exploit arguably quasi-random variations in future contestant strength due to the nature of the tournament draw. Second, we use an instrumental variable approach. For this, we utilize the structure of the contests and exploit unexpected wins of lower-ability contestants (upsets).\par 

In the tournaments, the individual contests are scheduled one after the other, so some contests are held before the contest that determines the next opponent, and others are held after. Consequently, some contestants already know which opponent they would meet in the next stage, while others do not know for sure at this point. The instrument we use is whether the opponent in the next stage is already known at the time of one's (current) contest. Thus, we instrument the expected ability of the future opponent by a binary variable indicating if the future opponent is known (=1) or not (=0). The order of contests in a tournament is scheduled randomly, ensuring the exogeneity of the instrument.
The relevance of the instrument comes from the realization of the opponent in the next stage. This meets the expectations if the higher-ability contestant of the parallel contest proceeds to the next stage but deviates from the expectations whenever the higher-ability contestant is (surprisingly) eliminated. Therefore, when the lower-ability contestant upsets the higher-ability contestant, the instrument shifts the expected ability of the opponent in the next stage when the schedule is such that the opponent is known at the time of one's (current) contest.\footnote{The complier population are those that would meet an opponent with lower than expected ability and get to know this before their own contest. There are, by construction, no defiers since they can not meet an opponent with higher ability than the higher-ability contestant in the parallel contest. This ensures that the monotonicity assumption is fulfilled.} \par

\section{Results}\label{sec:results}

In this section, we present the results of our empirical analysis. Section \ref{subsec:cont_het} shows how skill disparities between contestants affect individual performance and contest outcomes. In Section \ref{subsec:ect}, we show how a head start for the lower-ability contestant affects the individual performance of all contestants. In Section \ref{subsec:spill}, we investigate shadow effects across stages of a tournament.  \par

\subsection{Contestant Heterogeneity}\label{subsec:cont_het}
\subsubsection{Individual Performance}\label{subsubsec:within_contest}

We examine how the heterogeneity in contestant ability influences their performance. According to theory, increasing heterogeneity in ability among contestants reduces the motivation for the contestant with lower abilities to perform well, enabling the contestant with higher abilities to decrease performance. To investigate this, we distinguish between the higher-ability contestant (favorite) and the lower-ability contestant (underdog) for each contest and present the results separately.\par

The estimates for the lower-ability contestants are presented in Panel A of Table \ref{tab:table_main_perf}. Our results show that the performance of the lower-ability contestant decreases significantly due to the increasing heterogeneity of abilities among the contestants. The magnitude of the point estimates is quantitatively meaningful: an increase of one standard deviation in the contestants' ability ratio leads to a decrease of about 0.75 points in the 3-darts average of the lower-ability contestant.
Given a mean performance difference of 5 points between competing contestants, this effect is substantial and aligns with the theoretical predictions from the literature \cite{Lazear.1981, Rosen.1986}.\par

	\begin{table}[H]
		\centering
		\caption{Effect of ability ratio on individual performance} 
		\label{tab:table_main_perf}
		\begin{tabular}{l r r r r r r } 
			\toprule \toprule
			\textbf{}                    & ~~~~~~~ & (1)    & ~~~~~~~ &  (2)   & ~~~~~~~ &  (3)         \\ 
			\midrule
            \multicolumn{7}{l}{\textit{Panel A: Underdogs's performance}} \vspace{0.7em}\\ 
			Current ability ratio$^1$ ~~~~~~~~ & & -15.171*** & & -15.353*** & & -15.075***      \\
			                               & & (2.999)    & &  (3.013)   & &  (3.006)        \\
			Favorite world ranking         & &            & &  -1.537    & &  -1.078         \\
			                               & &            & &   (1.232)  & &  (1.257)        \\
			Underdog world ranking         & &            & & -3.410**   & &  -3.297**       \\
			                               & &            & &  (1.513)   & &  (1.547)        \\
			Favorite experience            & &            & &            & &   0.028**       \\
			                               & &            & &            & &  (0.013)        \\
			Underdog experience            & &            & &            & &  2.257**        \\
			                               & &            & &            & &  (1.049)        \\
			\midrule
            \multicolumn{7}{l}{\textit{Panel B: Favorite's performance}} \vspace{0.7em} \\ 
			Current ability ratio$^1$      & & 5.264**   & &  5.408**  & &  5.738**          \\
			                               & & (2.524)   & &  (2.443)  & &  (2.569)          \\
            Favorite world ranking         & &           & &  -1.016   & &  -0.920           \\
			                               & &           & &  (1.671)  & &  (1.672)          \\
			Underdog world ranking         & &           & &  -0.292   & &  -0.238           \\
			                               & &           & &  (0.988)  & &  (1.000)          \\
			Favorite experience            & &           & &           & &   2.373***        \\
			                               & &           & &           & &  (0.689)          \\
			Underdog experience            & &           & &           & &  0.008            \\
			                               & &           & &           & &  (0.012)          \\
            \midrule
			Stage FE                       & &  x         & &  x         & &  x              \\ 
            Tournament-by-Year FE          & &  x         & &  x         & &  x              \\
			Individual FE                  & &  x         & &  x         & &  x              \\
			All Covariates                 & &            & &            & &  x              \\
			\midrule 
            N                              & &   4776     & &   4776     & &   4776          \\ 
			\bottomrule \bottomrule
			\multicolumn{7}{l}{\footnotesize Notes: *, **, and *** represents statistical significance at the 10 \%, 5 \%, and 1 \%, respectively. \textit{All Covariates}}\\
			\multicolumn{7}{l}{\footnotesize ~~~~~~~~~    include information on who starts the contest, if the favorite performs at home, and if the underdog}\\
			\multicolumn{7}{l}{\footnotesize ~~~~~~~~~     performs at home. Full regression tables are available upon request from the authors. FE = fixed }\\
			\multicolumn{7}{l}{\footnotesize ~~~~~~~~~   effects. $^{1}$ Ratio is the favorite / underdog ability (measured as 3-darts average over the past 2 years).}\\
			\multicolumn{7}{l}{\footnotesize ~~~~~~~~~  Linear regression. Standard errors are clustered at the individual level.}
		\end{tabular}
	\end{table}

Contrary to the theory's predictions, increasing ability heterogeneity has opposing effects on the performance of higher-ability contestants. In Panel B of Table \ref{tab:table_main_perf}, we observe a small but statistically significant increase in the performance of higher-ability contestants with an increasing ability ratio among contestants. 
In sum, the effects of a one standard deviation increase in ability heterogeneity account for about 20\% of the between-contestants mean performance difference.\footnote{To ease the interpretation of the magnitude of the observed effects, we consider the skill levels of darts player Michael van Gerwen between 2009 and 2020. During this period, when he turned from an average to one of the best darts players in the world, he improved his 3-darts average from 94.3 to 100.4, or on average, by about 0.5 per year. Thus, the one standard deviation increase in ability heterogeneity is about the same as the improvements in van Gerwen's performance over two years.} \par

Importantly, the effects reported in Table \ref{tab:table_main_perf} remain consistent regardless of the treatment definition or the inclusion of further control variables in the analysis. As this is crucial for our identification strategy, we put particular emphasis on controlling for ability in different ways. We provide alternative specifications with different definitions of the outcome, treatment, and control variables in Section \ref{subsec:robustnes} (and more extensively in Appendix Section \ref{subsec:app_robust}).\par

One potential concern regarding our approach is that we treat each match as a separate contest.\footnote{We argue that each contest is a separate winner-takes-all contest, as the winner advances to the next round, which guarantees some prize money and the option to compete for even more, whereas the loser is eliminated from the tournament.} Nevertheless, the single contest is embedded into a multi-stage tournament, and tournament dynamics such as previous effort provided or the remaining opponents and prize money in play may influence player incentives and performance. To see whether our result is sensitive to tournament dynamics, we extended our analysis and included two sets of control variables. First, following \citeA{Brown.2014}, we include measures for the past effort provided by both contestants and the strength of the next opponent.\footnote{\citeA{Brown.2014} as well as \citeA{Lackner.2020} show that contestants exhibit forward-looking behavior and consider the strength of the upcoming component when deciding about how much to invest in the current stage. \citeA{Brown.2014} also show that past effort matters for performance in the current stage. For a careful analysis of contestants' forward-looking behavior, see Section \ref{subsec:spill}.} Second, we add detailed information regarding the prize money distribution. To control for past effort provided, we control for the number of legs both contestants played already in the tournament before the contest. To control for the difficulty of the upcoming stage, we control for the expected ability of the potential next opponent. The monetary incentive structure (distribution of prize money) is captured by the proportion of the tournament prize money that is at stake in the current contest and potential future contests. We summarize our results in Appendix Table \ref{apptab:table_main_perf_sens}. We conclude that the estimates are rather insensitive towards the inclusion of covariates controlling for tournament dynamics.\par

\begin{figure}[H]
\centering
      \caption{Individual performance - non-parametric estimates}
      \label{fig:9av_np}
    \begin{subfigure}{.48\textwidth}
      \centering
      \includegraphics[width=\textwidth]{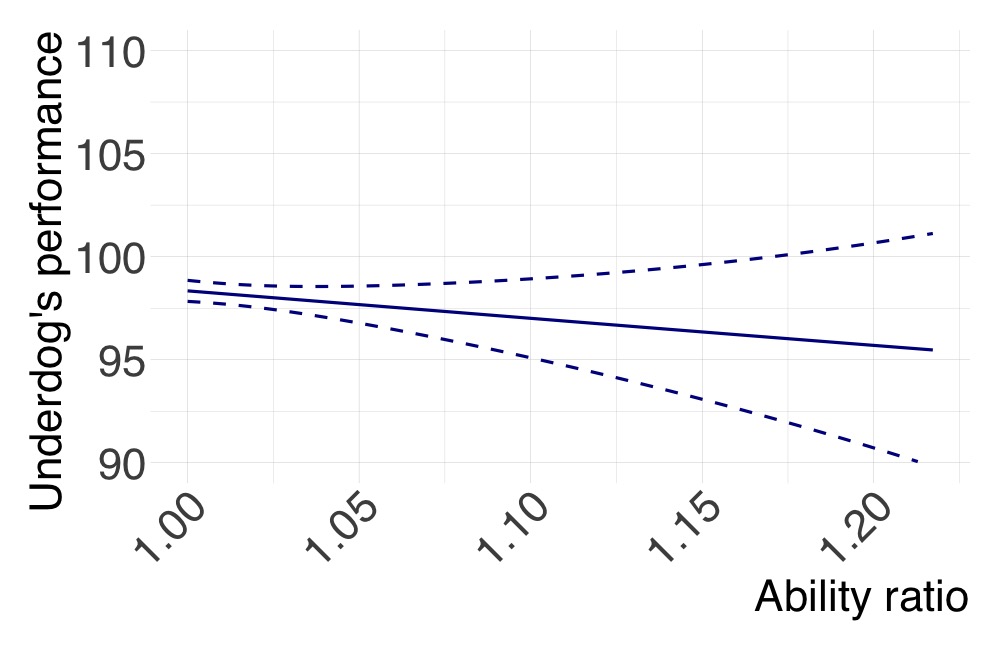}
      \caption{Lower-ability contestant performance}
      \label{fig:u9av_np}
    \end{subfigure}
    \hfill
    \begin{subfigure}{.48\textwidth}
      \centering
      \includegraphics[width=\textwidth]{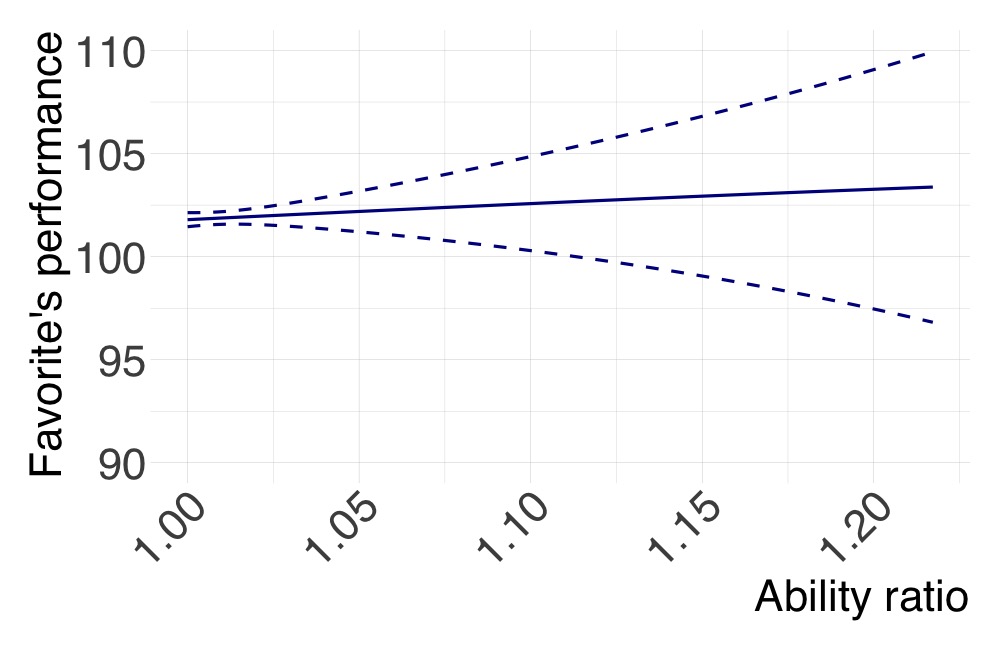}
      \caption{Higher-ability contestant performance}
      \label{fig:f9av_np}
    \end{subfigure}
        \begin{subfigure}{1\textwidth}
      \centering
      \includegraphics[width=.48\textwidth]{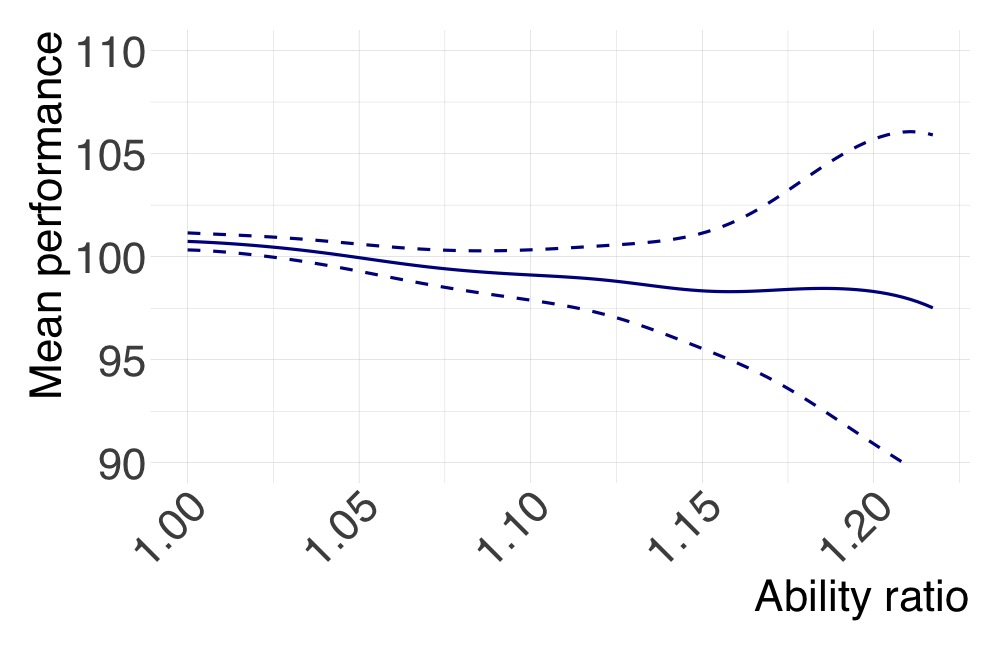}
      \caption{Mean performance}
      \label{fig:t9av_np}
    \end{subfigure}
	\footnotesize Notes: Non-parametric kernel regression. Based on Table \ref{tab:table_main_perf}, Column 3. Subsample without treatment outliers   \\
	\footnotesize ~~~~ (N=4728). The solid line shows the performance for different levels of the ability ratio. Dashed lines  \\
 	\footnotesize ~~~  show the 90 \% confidence intervals. ~~~~~~~~~~~~~~~~~~~~~~~~~~~~~~~~~~~~~~~~~~~~~~~~~~~~~~~~~~~~~~~~~~~~~~~~~~~~~~~~~~

\end{figure}

Another potential concern is the choice of a linear regression framework, i.e., that it is inflexible in detecting nonlinear effects and assumes a linear additive structure and a constant treatment effect. We employ a non-parametric kernel method from the causal machine learning literature to assess whether these concerns are real issues. The non-parametric estimates are presented in Figure \ref{fig:9av_np}.\footnote{We slightly adjust our set of covariates to implement the non-parametric estimation. In particular, we replace the individual fixed effect with the individual ability measures that describe the performance over the past two years. Note that in the robustness section, we show that exchanging these covariates does not substantially change the main results in the linear framework. We also use the prize money of each tournament as a control variable instead of the tournament-by-year fixed effect.} 
We plot the estimated average potential outcomes for each level of our treatment.
For two different treatment levels $A=a_1$ and $A=a_0$ (on the x-axis), the treatment effect can be calculated as $\theta(a_1,a_0) = \frac{E(Y(A=a_1))-E(Y(A=a_0))}{a_1-a_0}$. The intensity of the treatment in this example is $a_1-a_0$. The treatment level from which the treatment intensity is evaluated is $a_0$.\par

The resulting estimates provide two insights. First, they show that the treatment effect is relatively constant within our observed range of the treatment variable. We use this as evidence that the underlying assumption in the linear regression model of an average constant effect is reasonable in the investigated range of the treatment. Second, they quantitatively support the results reported in Table \ref{tab:table_main_perf}. Specifically, a one standard deviation increase in the treatment variable (0.05) decreases the potential outcomes by about 0.75 units for the lower-ability contestant and increases the potential outcomes for the higher-ability contestant by about 0.25 units.\par

So far, we have considered the effects of contestant ability heterogeneity on the individual performance of all contestants. Depending on the context, contest organizers may be interested in different measures, such as a high average performance or a balanced competition. Thus, a combination of both outcomes can be of interest depending on the contest designer's objective function. To provide a comprehensive view,  we also present effects on the average performance of both contestants in Figure \ref{fig:t9av_np}.\footnote{Since we observe the performance of both contestants separately, we can, in principle, measure performance in any (linear) combination of the two observed performances. For simplicity, we decided to focus on the three most prominent cases. The point estimate from estimating the effect of heterogeneity in contestant ability on aggregate performance can be obtained in Panel C of Table \ref{tab:table_base_1}, Column 1.} We find that increasing heterogeneity has a moderate negative effect on performance.\par

Throughout a contest, the incentives to perform might change depending on the course of the contest. Our measure of heterogeneity is defined before the contest begins. Thus, the first half of the contest, in particular, should be informative about immediate responses to contestant ability heterogeneity. In contrast, match dynamics may also affect effects in the second half. To provide a more detailed analysis, we divide each contest into two halves based on the realized number of legs played to obtain balanced samples.\par

Table \ref{tab:table_base_1} presents the results. In Column 1, for reference, we show the main result from the previous estimations of Table \ref{tab:table_main_perf}. Column 2 displays the performance in the contest's first half, while Column 3 provides the results for the second half. In Panel A, we observe a decrease in performance for the lower-ability contestants due to increasing contestant ability heterogeneity throughout both halves of the contest. The point estimates for both halves are close to the point estimates of the overall effect.\par

However, we observe meaningful differences for the higher-ability contestant in Panel B. In the contest's first half (Column 2), we find a notable positive effect with a point estimate more than twice as large as the overall effect (Column 1). In the second half (Column 3), however, we observe a small negative effect that is not statistically different from zero. The effect size in the first half is quantitatively relatively comparable for both higher-ability and lower-ability contestants. This finding strengthens our argument that increasing heterogeneity among contestants has contrasting effects on the performance of higher-ability and lower-ability contestants.\par

\begin{table}[H]
		\centering
		\caption{Results for different parts of the contest} 
		\label{tab:table_base_1}
		\begin{tabular}{l r r r r r } 
			\toprule
			\toprule
			& \multicolumn{1}{c}{\textbf{Performance}} & ~~~ &  \multicolumn{3}{c}{\textbf{Performance in }} \\
   			& \multicolumn{1}{c}{\textbf{within contest}} & ~~~ &  \multicolumn{1}{c}{\textbf{~~first half~~}} & ~ & \multicolumn{1}{c}{\textbf{second half}} \\
            \cline{2-2} \cline{4-6} \rule{0pt}{3ex}   
			        & (1)        & ~ & (2)   & ~ & (3)     \\ 


			\midrule
   			\multicolumn{6}{l}{\textit{Panel A: Underdog's performance}}  \\
            \rule{0pt}{0.01ex}\\
			Ability ratio            & -15.075*** & & -13.349*** & & -17.198***  \\
			                         & (3.006)    & & (4.088)    & & (3.939)     \\
            \rule{0pt}{0.3ex}\\
			\multicolumn{6}{l}{\textit{Panel B: Favorite's performance}}      \\
            \rule{0pt}{0.01ex}\\
			Ability ratio            & 5.738**   & & 12.303***   & & -2.035   \\
			                         & (2.569)   & & (3.492)     & & (3.541)  \\
            \rule{0pt}{0.3ex}\\

			\multicolumn{6}{l}{\textit{Panel C: Mean (combined) performance}}   \\
            \rule{0pt}{0.01ex}\\
			Ability ratio            & -4.506     & & 0.359      & & -10.168***  \\
			                         & (2.790)    & & (3.658)    & & (3.872)     \\
            \rule{0pt}{0.3ex}\\
			\midrule
			Stage FE ~~ & x  & & x & & x  \\ 
			Tournament-by-Year FE    & x  & & x & & x  \\ 
			Individual FE            & x  & & x & & x  \\ 
			All Covariates           & x  & & x & & x  \\
			\midrule 
			N                        &  4776    & &  4776   & &   4776   \\ 
			\bottomrule
			\bottomrule
			\multicolumn{6}{l}{\footnotesize Notes: *, **, and *** represents statistical significance at the 10 \%, 5 \%, and 1 \%, respectively. }\\
			\multicolumn{6}{l}{\footnotesize ~~~~~~~~~ Linear Regression. Standard errors are clustered at the individual level. Full regressions}\\
			\multicolumn{6}{l}{\footnotesize ~~~~~~~~~ can be found in Tables \ref{tab:table_main_perf} and \ref{tab:table_app_half}. FE = fixed effects.}
		\end{tabular}
	\end{table}

In Panel C of Table \ref{tab:table_base_1}, the effects on the average performance of both contestants are in line with the individual performance results. We observe a negative effect towards the end of the contest, while in the first half of the contest, the individual effects offset each other, and, on average, no performance decrease is observed. \par

\subsubsection{Subsample analyses} \label{subsubsec:het}
\begin{figure}[H]

      \caption{Subsample effects}
      \label{fig:het}
    \begin{subfigure}{.5\textwidth}
      \centering
      \includegraphics[width=\textwidth]{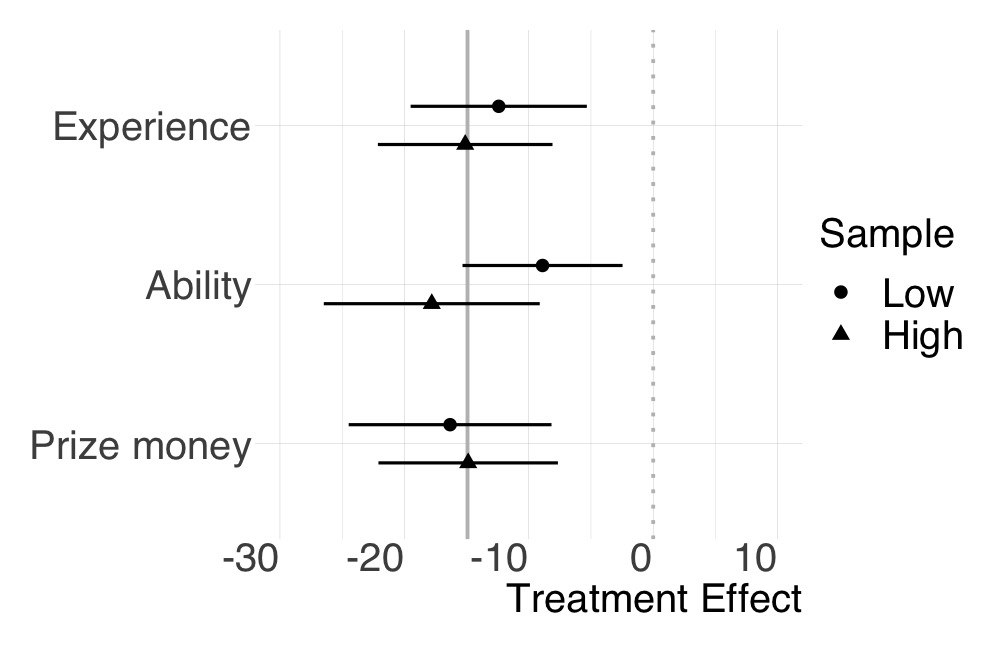}
      \caption{Lower-ability contestant performance}
      \label{fig:het_und}
    \end{subfigure}
    \hfill
        \begin{subfigure}{.5\textwidth}
      \centering
      \includegraphics[width=\textwidth]{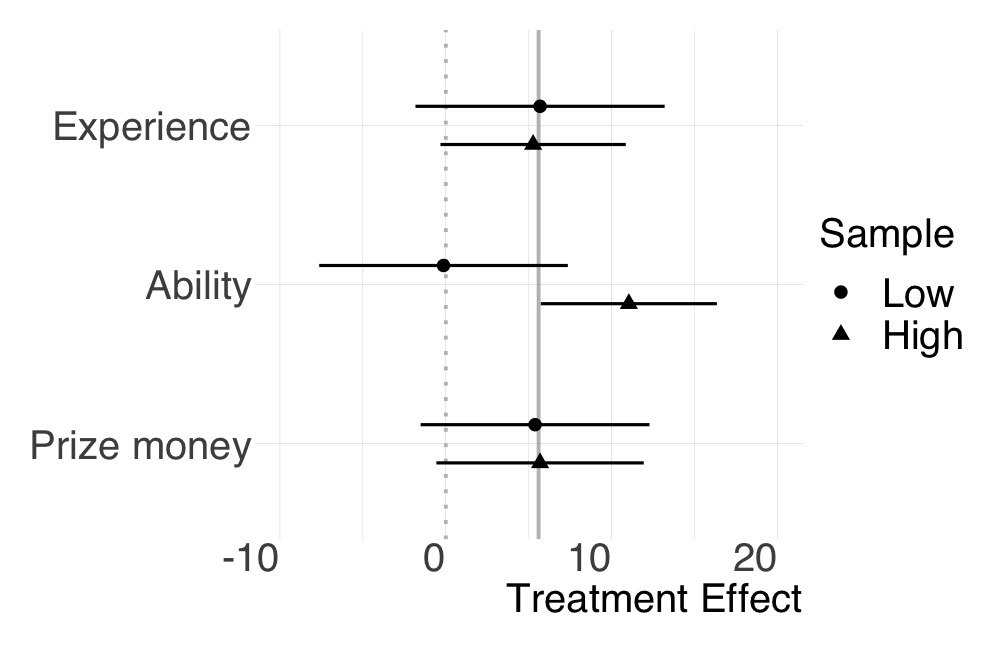}
      \caption{Higher-ability contestant performance}
      \label{fig:het_fav}
    \end{subfigure}
    \center{
	\footnotesize Notes: Linear regression on subsamples for low, i.e., below median, (circle) and high, i.e., above median,   \\
	\footnotesize ~~~~~~~~~~~ (triangle) values of the three heterogeneity variables. Specifications based on Table \ref{tab:table_main_perf}, Column 3.    ~~ \\
	\footnotesize ~~~~~~~~~~~ The grey vertical lines indicate the average treatment effect. Whiskers mark the 90 \% confidence  ~~\\
 	\footnotesize ~~~~~~  intervals.  \textit{Experience} is measured by years of playing darts. \textit{Ability} by the world ranking. ~~~~~ \\
   	\footnotesize ~~~~~~~ \textit{Prize money} is the total tournament purse, standardized to account for increasing prize money  \\ 
    \footnotesize ~ over the years. For more details, see Section \ref{sec:data}). ~~~~~~~~~~~~~~~~~~~~~~~~~~~~~~~~~~~~~~~~~~~~~~~~~~~~~~~~ \\}
\end{figure}

In this subsection, we explore whether the effect of contestant heterogeneity varies for different subpopulations defined by contestant and tournament characteristics. The aim is to investigate if specific subpopulations react differently to increasing ability heterogeneity. We focus on three characteristics that have meaningful interpretations in the tournament literature and have practical implications beyond the sports setting.  For this, we split the total sample by the contestant's experience, measured by the number of years the individual plays darts, the contestant's ability, measured by the position in the world ranking, and the total tournament purse (\textit{prize money}) distributed, signifying the importance of the tournament.
Samples are divided at the median value of each variable, and the subsamples are defined as \textit{High} (containing individuals with above median values in the respective characteristic) and \textit{Low} (containing individuals with below median values in the respective characteristic). The point estimates along with 90\% confidence intervals are shown in Figures \ref{fig:het_und} and \ref{fig:het_fav} for our two individual performance measures.\par

In Figures \ref{fig:het_und} and \ref{fig:het_fav}, we do not observe significant differences between the above (\textit{High}) and below (\textit{Low}) median groups; the estimates are close to the average estimate (which is depicted by the vertical solid lines). However, we do gain an interesting insight for the higher-ability contestants. We observe a significant positive effect of contestant ability heterogeneity only for high performers. In certain contests, such as innovation contests, the contest designer is particularly interested in the performance of the most talented contestants. This finding suggests that the most talented contestants, in particular, improve their performance as contestant heterogeneity increases. \par

\subsubsection{Robustness Checks} \label{subsec:robustnes}

In Section \ref{subsubsec:within_contest}, we showed that the effect of ability heterogeneity on performance differs when focusing on higher-ability or lower-ability contestants.
In Appendix \ref{subsec:app_robust}, we conduct a battery of alternative specifications to test the robustness of our result.
We show that our main findings are virtually unaffected by several robustness checks, including the definition of the treatment variable (Columns 1 and 2, Table \ref{apptab:table_rob_new}), the sample definition (Columns 3 and 4, Table \ref{apptab:table_rob_new}), and a more extensive set of control variables (Columns 5 and 6, Table \ref{apptab:table_rob_new}). Particularly, we show different ways to proxy for the individual abilities of the contestants. Furthermore, we show that our result is robust to an alternative definition of the outcome variable and a differential set of outcome variables (Table \ref{apptab:table_rob_new}, Column 7 and Table \ref{apptab:table_base_1_180}). We provide a more detailed description of each robustness check in Appendix \ref{subsec:app_robust}. Overall, we conclude that our main results are robust to several alternative specifications and provide a good deal of evidence that skill disparities among contestants have asymmetric effects on the performance of contestants. \par

\subsubsection{Contest Outcomes}\label{subsubsec:contest_outcomes}

So far, we have shown how heterogeneity among contestants affects individual performance. In this section, we complement our analysis and investigate the effect of ability heterogeneity on contest outcomes instead of individual performance. Consistent with previous studies, such as \citeA{Brown.2014}, we use a binary variable indicating whether the higher-ability contestant won the contest as one of our outcome variables. In addition, we use the realized length of the contest and the number of top performances, i.e., scoring a 180 -- a common statistic in professional darts -- to approximate the attractiveness of contests.\par

Table \ref{tab:table_base_0} summarizes our findings. We observe that increasing heterogeneity in contestant ability leads to a decrease in the contest length (Panel B), i.e., contests in which the decision as to who is the winner is made more quickly.
Similarly, the number of top performances (Panel C) decreases, indicating a less interesting contest for spectators. The higher-ability contestant becomes more likely to win in the respective contests as the ability ratio between the contestants increases (Panel A). At this point, it is important to note that part of the effect on the contest outcome (e.g., whether the favorite wins) is \textit{mechanical}, as by increasing the heterogeneity between contestants, the winning probability of the higher-ability contestant increases even without any change in the individual performance by either of the contestants.\par 

\begin{minipage}{0.47\textwidth}
	\begin{table}[H]
		\centering
		\caption{Results for contest outcomes} 
		\label{tab:table_base_0}
		\begin{tabular}{l r r} 
			\toprule \toprule
			                     & \multicolumn{1}{c}{(1)}     &  \multicolumn{1}{c}{(2)}              \\ 
			\midrule
   			\multicolumn{3}{l}{\textit{Panel A: Higher-ability contestant win}}  \\[0.35\normalbaselineskip]
			Ability ratio        & 1.420*** & 1.423***          \\
			                     & (0.156)  & (0.159)           \\ [0.5\normalbaselineskip]
   			\multicolumn{3}{l}{\textit{Panel B: Contest length}}     \\ [0.35\normalbaselineskip]
			Ability ratio        & -0.347*** & -0.346***          \\
			                     & (0.053)  & (0.053)           \\ [0.5\normalbaselineskip]
   			\multicolumn{3}{l}{\textit{Panel C: Number max. performance}}     \\ [0.35\normalbaselineskip]
			Ability ratio        & -5.665*** & -5.037***          \\
			                     & (1.786)  & (1.697)           \\ [0.5\normalbaselineskip]
			\midrule
			All FE    & x & x  \\
			All Covariates       &   & x  \\
			\midrule 
			N                    &  4776    & 4776      \\ 
			\bottomrule \bottomrule
			\multicolumn{3}{l}{\footnotesize Notes: Linear regression. *** indicates statistical   }\\
			\multicolumn{3}{l}{\footnotesize ~~~~~~~~~ significance at the 1 \% level. Full   }\\
			\multicolumn{3}{l}{\footnotesize ~~~~~~~~~  regressions can be found in Table \ref{tab:table_app_contest}. }
		\end{tabular}
	\end{table}
\end{minipage} 
\begin{minipage}{0.52\textwidth}
    \begin{figure}[H]
		\centering
		\caption{Contest outcomes and betting market data}
		\label{fig:feed_desc}
			\includegraphics[width=1\textwidth]{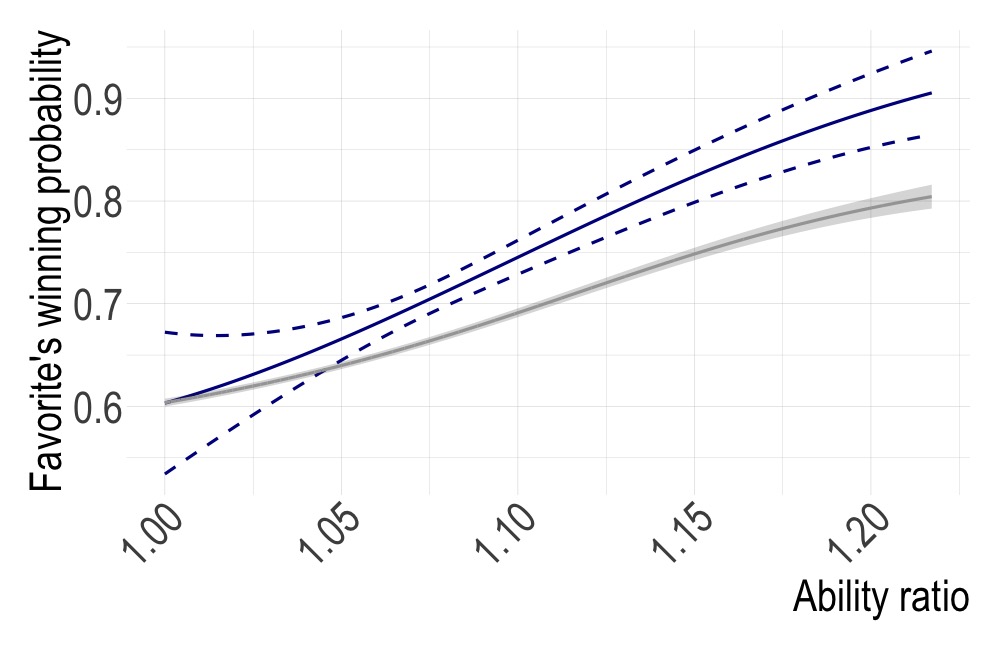}\\ [0.29\normalbaselineskip]
        \footnotesize Notes: Non-parametric kernel regression. Based on \\ [0.29\normalbaselineskip]
		\footnotesize ~~~~~~ Table \ref{tab:table_app_contest}, Column 3. Subsample without  \\ [0.29\normalbaselineskip]
		\footnotesize ~~~~~~~~~ treatment outliers (N=4728). The blue line \\ [0.29\normalbaselineskip]
  		\footnotesize ~~~~~  shows the expected favorite winning   ~~~~~  \\ [0.29\normalbaselineskip]
    	\footnotesize ~~~~~~~~~~  probability for different levels of the ability     ~   \\ [0.29\normalbaselineskip]
    	\footnotesize ~~~~ ratio. Dashed lines show the 90 \%     ~~~~~~~ \\[0.29\normalbaselineskip]
    	\footnotesize ~~~~~~~~~~~~ confidence intervals. The grey line shows the  ~ \\[0.29\normalbaselineskip]
    	\footnotesize ~~~ respective winning prob. implied by ~~~~ \\ [0.29\normalbaselineskip]
        \footnotesize ~~ bookmakers' betting odds. ~~~~~~~~~~~~~~
	\end{figure}
\end{minipage}
~\par

To visualize the individual performance responses and their impact on the contest outcome, we benchmark our estimated effects on the winning probability of the higher-ability contestant against the predicted winning probabilities of the higher-ability contestant derived from bookmakers' betting odds for different levels of ability heterogeneity. To accomplish this, we employ the same non-parametric kernel method described earlier to estimate the effect of contestant ability heterogeneity on the winning probability of the higher-ability contestant. This is illustrated by the blue line in Figure \ref{fig:feed_desc}. We observe that as contestant heterogeneity increases, the winning probability of the higher-ability contestant also increases, reaching a winning probability of approximately 90\%. 
In addition, the bookmakers' predicted win probabilities are shown in gray. For this purpose, the betting odds of each contest were converted into the implied win probabilities and smoothed using a local kernel estimation for the corresponding ability ratio. Interestingly, we find that the observed increase in the winning probability of the higher-ability contestant is substantially larger than what is implied by the bookmakers' predictions. This suggests that the observed results regarding the probability of the higher-ability contestant winning are not solely driven by the initial ability differences in ex-ante winning probabilities but also by the individual performance responses to the ability heterogeneity of contestants within the contest.\footnote{To complement the analysis on contest outcomes, we show in Figure \ref{fig:het_favw} that the observed increase in performance among the top contestants (see Chapter \ref{subsubsec:het}) also translates into an increased probability of winning the contest for top contestants.}\par

\subsection{Head start}\label{subsec:ect}

A common approach to mitigate the adverse incentive effects resulting from contestant heterogeneity is to implement interventions that bias the contest success function, thereby reducing the effective level of heterogeneity among contestants. Affirmative action policies serve as prominent examples of such interventions \cite{Chowdhury.2023, Schildberg.2023}.\par

Based on the identification strategy discussed in Section \ref{subsec:ident_hs}, our setting allows us to empirically test how biased contests, in which one contestant is given an advantage, affect the performance of both higher-ability and lower-ability contestants. Additionally, we investigate whether the effect of a head start varies for different levels of contestant heterogeneity. We use a special feature of darts contests to study how an initial advantage for one contestant affects individual performance. Starting a contest in darts offers a built-in advantage because the starting contestant never has fewer but sometimes more moves than the opponent. With the starting right changing at each leg and an odd maximum number of legs, the advantage transmits throughout the contest \cite{Goller.2023}.\par

In Column 1 of Panel A of Table \ref{tab:table_main_head}, we present our findings regarding the impact of a head start on the contest outcome. We observe that a head start substantially affects the contest outcome, significantly influencing the likelihood of the higher-ability contestant winning. Specifically, when the lower-ability contestant is given a head start, the higher-ability contestant becomes approximately 7 percentage points less likely to win the contest.\par

	\begin{table}[H]
		\centering
		\caption{Effect of a head start for the lower-ability contestant} 
		\label{tab:table_main_head}
		\begin{tabular}{l r r r r r r r} 
			\toprule
            \toprule
   			& \multicolumn{1}{c}{\textbf{(1)}} &  &  \multicolumn{1}{c}{\textbf{(2)}} & & \multicolumn{1}{c}{\textbf{(3)}} & & \multicolumn{1}{c}{\textbf{(4)}}\\
      \midrule
			& \multicolumn{1}{c}{\textbf{~~~Favorite ~~~}} & ~~ &  \multicolumn{5}{c}{\textbf{Performance}}  \\
            \cline{4-8}
   			& \multicolumn{1}{c}{\textbf{win}} &  &  \multicolumn{1}{c}{\textbf{Favorite}} & & \multicolumn{1}{c}{\textbf{Underdog}} & & \multicolumn{1}{c}{\textbf{Average}}\\
            \cline{2-2} \cline{4-4} \cline{6-6} \cline{8-8} \\  
   			\multicolumn{8}{l}{\textit{Panel A: Baseline effect}}    \\            \vspace{0.01em}   \\
			Underdog head start  ~~~          & -0.075***    & & -0.228    & & 0.688***  & & 0.277    \\
			                                 & (0.017)      & & (0.280)   & & (0.202)   & & (0.169)  \\
            \rule{0pt}{0.3ex}\\
   			\multicolumn{8}{l}{\textit{Panel B: Interaction effect}}    \\            \vspace{0.01em}\\
			Ability ratio                    & 1.283***     & & 5.923**   & & -16.380*** & & -4.208  \\
			                                 & (0.178)      & & (2.490)   & & (3.629)    & & (2.960) \\
			Underdog head start  ~~~          &              & &           & &            & &         \\
			x low ability ratio$^1$ ~~~          & -0.082***    & &  -0.095   & & 0.448      & & 0.309   \\
			                                 & (0.026)      & & (0.363)   & & (0.336)    & & (0.257) \\
			x medium ability ratio$^1$ ~~~       & -0.100***    & &  -0.390   & & 0.794**    & & 0.302   \\
			                                 & (0.023)      & & (0.369)   & & (0.332)    & & (0.245) \\
			x high ability ratio$^1$ ~~~         & -0.045**     & &  -0.208   & & 0.840**    & & 0.219   \\
			                                 & (0.020)      & & (0.376)   & & (0.376)    & & (0.267) \\
            \rule{0pt}{0.3ex}\\
            \midrule
			Stage FE  & x  & & x & & x & & x  \\ 
			Tournament-by-Year FE    & x  & & x & & x & & x  \\ 
			Individual FE            & x  & & x & & x & & x  \\ 
			All Covariates           & x  & & x & & x & & x  \\
            \midrule
            N                        & 4776 & & 4776 & & 4776 && 4776  \\ 		 
			\bottomrule
            \bottomrule
			\multicolumn{8}{l}{\footnotesize Notes: *, **, and *** represents statistical significance at the 10 \%, 5 \%, and 1 \%, respectively. Standard}\\
			\multicolumn{8}{l}{\footnotesize ~~~~~~~~~  errors are clustered at the individual level. Control variables in Panel A  as in Table \ref{tab:table_main_perf}, Column 3.}\\
			\multicolumn{8}{l}{\footnotesize ~~~~~~~~~     Full results can be found in the Appendix Table \ref{tab:table_app_head}. FE = fixed effect. $^1$ “low” refers to ability}\\
            \multicolumn{8}{l}{\footnotesize ~~~~~~~~~  ratios below 1.026, “high” to ability ratios above 1.067 and “medium” to those in between.}
   
		\end{tabular}
	\end{table}

In Columns 2 to 4 of Panel A, we show this is not a purely mechanical effect resulting from changes in winning probabilities. Instead, it also has an impact on the individual responses of contestants. Our analysis reveals that lower-ability contestants experience a significant improvement in their performance when given a head start. This aligns with the argument that a head start for the lower-ability contestant serves as a method to "level the playing field" by increasing the motivation for the lower-ability contestant. We find no evidence that the performance of higher-ability contestants is significantly affected when the lower-ability contestant is given a head start. In line with our main results, the point estimate for the higher-ability contestant is negative (opposite sign compared to the lower-ability contestant) but not statistically different from zero. Also note that in terms of effect sizes, the average effect of a head start roughly compensates for the average effect of a one s.d. increase in contestant heterogeneity for both contestants.\par

In Panel B of Table \ref{tab:table_main_head}, we investigate whether the importance of a head start for the lower-ability contestant varies with the heterogeneity in the ability among contestants. To investigate this, we interact the head start variable with an indicator variable for a low, medium, or high degree of contestant heterogeneity.\footnote{We construct the variable for the degree of contestant heterogeneity by dividing the variable for the ability ratio into three equally sized parts, i.e., “low” refers to ability ratios below 1.026, “high” to ability ratios above 1.067 and “medium” to those in between.} We find that the head start matters more when heterogeneity among contestants is large.\par

These findings suggest that biasing a contest towards lower-ability contestants can efficiently increase incentives and improve individual performance. The contest becomes more competitive by providing a head start to the lower-ability contestant. It leads to enhanced performance from the lower-ability contestants without meaningful negative effects on the performance of the higher-ability contestants.\par

\subsection{Shadow effects} \label{subsec:spill}

Many tournaments do not consist of single-stage contests where participants compete solely for immediate prizes. Instead, innovation tournaments and public tenders, for example, often involve multiple stages in which contestants compete for immediate rewards and the opportunity to win additional prizes in future stages. We explore whether contestants in multi-stage tournaments exhibit forward-looking behavior. In particular, we investigate how the expected strength of future competitors affects the behavior of contestants in the current contest in dynamic tournament scenarios.\par

Anecdotal evidence suggests that professionals often focus solely on the current stage of a tournament and do not consider future opponents when making decisions about their level of effort. However, theoretical considerations and empirical evidence suggest that in multi-stage tournaments, future opponents may have shadow effects on the current stage \cite{Lackner.2020}. According to theory, stronger future opponents can decrease the consolidation value for both contestants in the current stage, leading to decreased effort from both participants. However, the higher-ability contestant is likely to experience a more significant decrease in performance compared to the lower-ability contestant. As a result, the probability of the higher-ability contestant winning the contest is expected to be lower due to this differential impact on performance \cite{Brown.2014}.\par

\begin{table}[H]
		\centering
		\caption{Shadow effects in multi-stage tournaments} 
		\label{tab:table_main_fut}
		\begin{tabular}{l r r r r r r r} 
			\toprule
            \toprule
               			& \multicolumn{1}{c}{\textbf{(1)}} &  &  \multicolumn{1}{c}{\textbf{(2)}} & & \multicolumn{1}{c}{\textbf{(3)}} & & \multicolumn{1}{c}{\textbf{(4)}}\\
      \midrule
			& \multicolumn{1}{c}{\textbf{~~~Favorite ~~~}} & ~ &  \multicolumn{5}{c}{\textbf{Performance}}  \\
            \cline{4-8}
   			& \multicolumn{1}{c}{\textbf{win}} &  &  \multicolumn{1}{c}{\textbf{Favorite}} & & \multicolumn{1}{c}{\textbf{Underdog}} & & \multicolumn{1}{c}{\textbf{Average}}\\
            \cline{2-2} \cline{4-4} \cline{6-6} \cline{8-8} \vspace{0.7em}\\   
			\multicolumn{8}{l}{\textit{Panel A: Linear regression}}      \\
            \rule{0pt}{0.01ex}\\
			Expected ability$^1$  ~~~~~~~ & -0.004* & & -0.063*   & & 0.066   & & -0.019  \\
			(next opponent)               & (0.002) & & (0.032)   & & (0.043) & & (0.028) \\
            \rule{0pt}{0.3ex}\\
			\multicolumn{8}{l}{\textit{Panel B: Instrumental variable}}  \\
            \rule{0pt}{0.01ex}\\
			Expected ability$^1$          & -0.018** & & -0.563*** & & 0.194   & & -0.184    \\
			(next opponent)               & (0.007)  & & (0.169)   & & (0.128) & &  (0.115)  \\
            \rule{0pt}{0.3ex}\\ 
            \midrule
            N & 4448 & & 4448 & &4448 & & 4448 \\
			\bottomrule
            \bottomrule
			\multicolumn{8}{l}{\footnotesize Notes: *, **, and *** represents statistical significance at the 10 \%, 5 \%, and 1 \%, respectively. }\\
			\multicolumn{8}{l}{\footnotesize ~~~~~~~~~ Standard errors are clustered at the individual level. Control variables in Panel A  as in Table}\\
			\multicolumn{8}{l}{\footnotesize ~~~~~~~~~  \ref{tab:table_main_perf}, Column 3. Panel B uses tournament-year and stage fixed effects. Full results can be found}\\
   			\multicolumn{8}{l}{\footnotesize ~~~~~~~~~    in Appendix Table \ref{tab:table_app_fut} and first-stage estimates in Table \ref{tab:table_app_2sls}. Panel B is based on a first-stage   }\\
         	\multicolumn{8}{l}{\footnotesize ~~~~~~~~~ F-Test of 346.7. $^1$ If the next opponent is known, this is their ability; otherwise, it is the ability }\\
         	\multicolumn{8}{l}{\footnotesize ~~~~~~~~~ of the favorite of the contest that will determine the next opponent.}
		\end{tabular}
	\end{table}

 Table \ref{tab:table_main_fut} presents the results from our two estimation approaches. Column 1 of Panel A shows results consistent with previous findings by \citeA{Brown.2014}. We observe that stronger future opponents have a negative effect on the probability of the higher-ability contestant winning the current contest. This result confirms the notion that the presence of stronger future opponents influences current performance outcomes.\footnote{In Figure \ref{fig:app_fwin_future}, we also provide non-parametric estimates on this result, which are consistent with the linear estimates.}\par

In Columns 2 to 4, we provide new insights into the underlying mechanisms driving this result. Specifically, we find that the higher-ability contestant exhibits a decrease in performance when faced with a stronger future opponent. However, we do not find a negative response from the low-ability contestant. While this finding aligns with the theoretical predictions regarding the relative decrease in performance, it does not fully confirm the theoretical expectations of an absolute decrease in the performance of both contestants. In Panel B, we show results using a different identification approach and exploit changes in the strength of future opponents due to upsets, which occur before the current match begins, as a source of exogenous variation. The results confirm the findings of Panel A.\footnote{Importantly, we show that the result in this section does not affect the result of Section \ref{subsubsec:within_contest}. To show this, we add control variables for the strength of the upcoming opponent to our primary regression model in Table \ref{apptab:table_main_perf_sens}. The main results are unaffected.} For the interpretation of the (larger) effect sizes, it is important to note that this is a treatment effect for the complier population, i.e., those individuals who know before their own contest that they would face an opponent with lower-than-expected ability in the next stage.\par

\section{Mechanisms}\label{sec:disc}

In Section \ref{subsec:cont_het}, we demonstrated that our findings regarding the lower-ability contestant align with the conventional wisdom that contestant heterogeneity can deter both effort provision and performance. However, this does not hold for the higher-ability contestant. This section discusses the potential of strategic choices and different behavioral approaches to explain our observed result.\footnote{A primary question to consider is whether effort influences performance in our context. In darts, physical effort is less integral than in other sports. However, there is still a clear connection between effort and performance,, encompassing mental preparedness, concentration, and focus maintenance. Results for the lower-ability contestant strengthen the reasoning that there is a link. We thus assume a monotone relationship between effort and performance and, in line with the tournament literature, model effort instead of performance in this section.}\par

\subsection{Theory-guided mechanism discussion}\label{subsec:tdisc}
 
We begin with the simple two-player Tullock contest that is usually used to motivate the common wisdom of a negative incentive effect of contestant heterogeneity on both contestants. Two contestants, $l$ and $h$, compete for a reward $R_i$, which might be differently evaluated as $R_l$ and $R_h$, by choosing the effort level $e_i$, i.e., $e_l$ and $e_h$. We assume that contestant $l$ is the lower-ability contestant while contestant $h$ is the higher-ability contestant. To characterize contestants' ability, we define a parameter $\theta_i$. For simplicity, we normalize the ability of the lower-ability contestant $\theta_l$ to 1. Consequently, the higher-ability contestants' ability $\theta_h$ with $\theta_h \geq 1$ then equals their relative skill advantage, as $\frac{\theta_h}{\theta_l}=\theta_h$. To ease notation, in the following $\theta_h$ denotes the relative skill advantage of the higher-ability contestant, i.e., the contestant heterogeneity. This parameter has the advantage of having a similar interpretation as our treatment variable. We assume that the contestant's contest success function takes on a logit form so that a contestant i’s probability of winning the contest is $ p_i =\frac{\theta_i e_i}{ \sum_{i=l,h} \theta_i e_i}$.\par

Contestants $l$ and $h$ maximize the expected payoff, which is the difference between their expected reward and cost: $ \pi_i = p_i*R_i - e_i$. For simplicity, we assume linear effort costs. This yields the following first-order conditions for our contestants:

\begin{center}

Contestant $l$: $\frac{\theta_h e_h}{(e_l + \theta_h e_h)^2} R_l - 1 = 0$ \hspace{3 em} Contestant $h$: $\frac{\theta_h e_l}{(e_l + \theta_h e_h)^2} R_h - 1 = 0$,
    
\end{center}

\noindent
and the following equilibrium effort:

\begin{center}
    $e_l^* = \frac{\theta_h R_l^2R_h}{(R_l + \theta_hR_h)^2} $ and $e_h^* = \frac{\theta_h R_lR_h^2}{(R_l + \theta_hR_h)^2} $.
\end{center}

\begin{figure}[H]
  \caption{Model predictions}
  \label{fig:model}
  \begin{subfigure}{.45\textwidth}
    \centering
    \includegraphics[width=\textwidth]{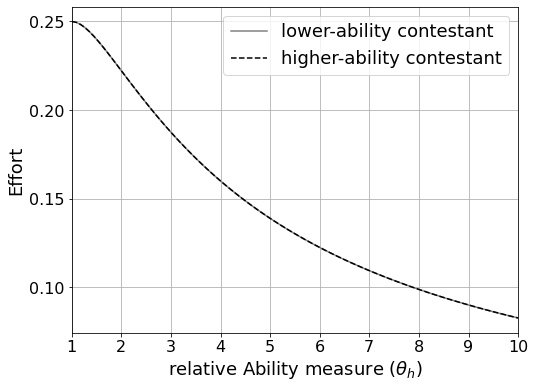}
    \caption{Baseline model}
    \label{subfig:model1}
  \end{subfigure}
  \hfill
  \begin{subfigure}{.45\textwidth}
    \centering
    \includegraphics[width=\textwidth]{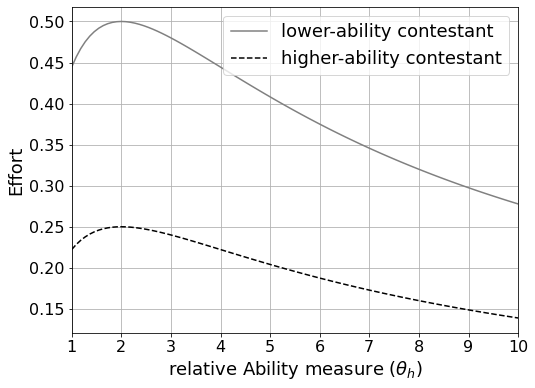}
    \caption{Differential evaluation of R: $R_l = 2 > R_h = 1$}
    \label{subfig:model2}
  \end{subfigure}

  \vspace{10pt}

  \begin{subfigure}{.45\textwidth}
    \centering
    \includegraphics[width=\textwidth]{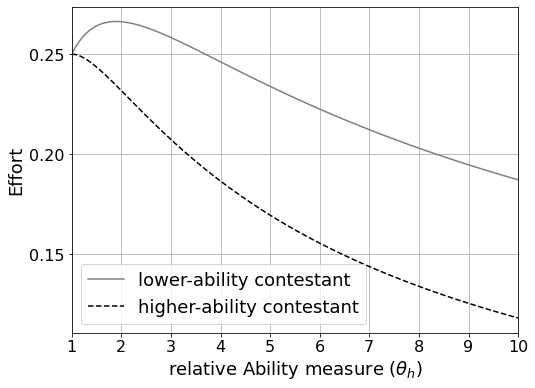}
    \caption{$R_l$ depends on heterogeneity}
    \label{subfig:model3}
  \end{subfigure}
  \hfill
  \begin{subfigure}{.45\textwidth}
    \centering
    \includegraphics[width=\textwidth]{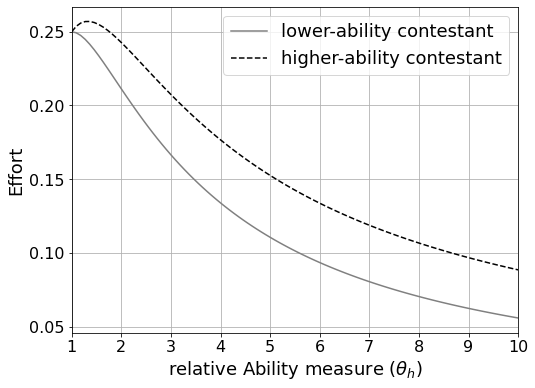}
    \caption{Choking of higher-ability contestant}
    \label{subfig:model4}
  \end{subfigure}
  \hfill
  \begin{center} 
    \begin{subfigure}{.45\textwidth}
      \centering
      \includegraphics[width=\textwidth]{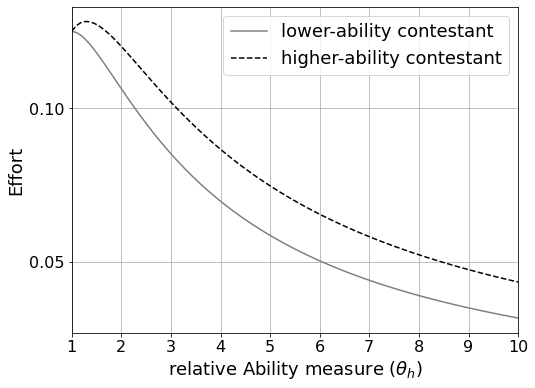}
      \caption{Anticipated regret of higher-ability contestant}
      \label{subfig:model5}
    \end{subfigure}
  \end{center}
      \center{
	\footnotesize Notes: Numerical solutions to the system of equations presented in this section. The vertical axis measures ~~~~~~~ \\
	\footnotesize ~~~~~~~~   the effort of the two contestants. The horizontal axis measures the relative ability advantage of higher- ~~ \\
 	\footnotesize ~~ ability contestant. The higher-ability contestant is $\theta_h$ times more skilled than the other contestant.}
\end{figure}

We illustrate the resulting equilibrium effort functions in Figure \ref{subfig:model1}, assuming the rewards for both contestants are identical ($R_l = R_h = 1$). Both contestants decrease effort provision with increasing contestant heterogeneity ($\theta_h$). \par

The "common wisdom" suggests that the lower-ability contestants reduce effort with increasing heterogeneity, and the higher-ability contestants reduce effort provision since they anticipate this behavior. A potential explanation for the non-equilibrium effort of the higher-ability contestant is, therefore, that the higher-ability contestants do not anticipate the lower-ability contestant's effort drop. Related to this argument, \citeA{Nagel.1995} argues that people exhibit different degrees of depth of reasoning.\footnote{A level-0 type exhibits nonstrategic behavior and adheres to a straightforward decision rule. A level-1 type behaves as if they best respond to the belief that the other individual is a level-0 type. This logic can be extended to various other types as well. \citeA{Bernard.2010} incorporates level-k type reasoning into contest theory and compares its predictions with those derived from standard Nash equilibrium. He shows that level-k players, relative to Nash equilibrium players, exert less effort at any given level of k.} While low degrees of strategic reasoning could explain constant effort of the higher-ability contestant, it does not align with their observed effort increase. This would suggest that higher-ability contestants mistakenly expect lower-ability contestants to increase effort with greater contestant heterogeneity.\par

Even though our empirical results do not support this idea, under what conditions might we expect the lower-ability contestant to increase effort as contestant heterogeneity increases? An argument could be made that $R$ differs for the contestants. While both contestants ostensibly compete for the same reward--advancing to the next stage--the evaluation of this reward may vary. The lower-ability contestant may derive an additional, non-monetary benefit from defeating a higher-ability contestant.\par

We illustrate the resulting equilibrium effort functions with two examples. In Figure \ref{subfig:model2}, we depict these functions when the lower-ability contestant's reward is $x$ times that of the higher-ability contestant.\footnote{In our numerical example, we show the case for $x = 2$; thus, the lower-ability contestants value winning the contest twice as much as the higher-ability contestants.} Under this assumption, the lower-ability contestant increases effort if $\theta_h < x$ but decreases effort otherwise. Since our observed values of $\theta_h$ may be smaller than $x$, erroneously anticipating that the lower-ability contestant gains an additional benefit from winning could explain our observed result. \par

How plausible is this explanation? If this explanation holds, we would anticipate the additional benefit received by the lower-ability contestant to depend on contestant heterogeneity ($\theta_h$) rather than being constantly higher. In Figure \ref{subfig:model3}, we depict the equilibrium effort functions when the lower-ability contestant's reward is $\theta_h^\alpha$ times that of the higher-ability contestant. In this scenario, the lower-ability contestant increases effort for small values of $\theta_h$ and decreases effort afterwards.\footnote{In our numerical example, we show the case for $\alpha = 0.2$.} The higher-ability contestant, on the other hand, reduces effort even for small values of $\theta_h$. Therefore, we conclude that while award asymmetries can account for the belief that the lower-ability contestant increases effort when contestant heterogeneity increases, they cannot explain the response from the higher-ability contestant, making out-of-equilibrium beliefs an unlikely reason for our results.\par

Another reason for the observed pattern could be the well-documented 'choking under pressure' phenomenon \cite{Baumeister.1984, Ariely.2009}. \citeA{Yerkes.1908} suggest an optimal arousal level for task performance, with deviations leading to decreased performance. \citeA{Camerer.2005} argues that increased incentives can involuntarily shift individuals from automatic to controlled mental processes, which is less effective for highly practiced tasks. Numerous empirical studies in different environments support this notion, revealing that individuals often underperform under pressure \cite{Dohmen.2008, Apesteguia.2010, Schlosser.2019}. \par

We argue that contestant homogeneity (i.e., low contestant heterogeneity) intensifies competitiveness in contests, creating pressure -- especially on the higher-ability contestant. We rationalize this by considering that the higher-ability contestant (the favorite) is expected to advance to the next stage. Our results in Section \ref{subsec:spill} support this idea. We demonstrate that the strength of the next opponent -- relevant only to the winner of the current contest -- negatively impacts the performance of the higher-ability contestant. In contrast, the lower-ability contestant does not exhibit this forward-looking behavior.
This suggests that the higher-ability contestant's expectations of advancement may lead to a response to competitive pressure (contestant homogeneity), potentially resulting in performance declines.\par

To model this behavior, we include a choking parameter in our model set-up making the higher-ability contestant's cost function dependent on $\theta_h$.\footnote{The idea of introducing a choking parameter in the cost function is related to an approach by \citeA{Bilen.2023}. The assumptions under which choking occurs differ from our setting.} In particular, we assume that the costs of the higher ability contestant are $\theta_h^{-\alpha} * e_h$. We derive the resulting equilibrium effort $e^*=(\frac{\theta_h^{\alpha + 1} R_hR_l^2}{(R_1 + \theta_h^{\alpha + 1}R_h)^2}, \frac{\theta_h^{2\alpha + 1} R_lR_h^2}{(R_l + \theta_h^{\alpha + 1}R_h)^2})$. While $\frac{\partial e_l}{\partial \theta_h}$ is strictly negative, $\frac{\partial e_h}{\partial \theta_h}$ is positive for low values of $\theta_h$, while negative for higher values of $\theta_h$. We illustrate a numerical example in Figure \ref{subfig:model4} for $\alpha = 0.2$.\footnote{The observed differential response to heterogeneity among contestants for lower-ability and higher-ability contestants is not sensitive to the choice of the parameter $\alpha$ for $\alpha > 0$.} As can be seen in Figure \ref{subfig:model4}, this model aligns with our empirical result from Section \ref{subsec:cont_het}, since we observe only values of $\theta_h$ between 1 and 1.5.\footnote{Note that our results in Section \ref{subsec:cont_het} should not be interpreted as evidence that any degree of heterogeneity among contestants is performance-enhancing for the higher-ability contestant. To underline this, remember that we show in Table \ref{tab:table_base_1} that in the second half, when winning probabilities are, on average, even more in favor of the higher-ability contestant, we find a small negative but statistically insignificant effect on the performance of the higher-ability contestant.} \par

An alternative explanation for the obtained empirical result may relate to the idea of anticipated regret \cite{Filiz.2007, filiz.2010, Hyndman.2012}. Even though the actual revenues for winning and losing are not opponent-specific (a win gets you higher prize money and the opportunity to compete for more, while a loss eliminates you from the tournament), there is the chance that the perception of the result is opponent-specific. A potential scenario could, therefore, be that the higher-ability contestant fears an upset (a loss against a substantially less skilled opponent) as this result would potentially affect his self-confidence or other factors despite not having different actual costs than a loss against an equally skilled opponent. To model this, we have to introduce (non-zero) returns to losses ($T_i$) to our model, which gives the following payoff functions:

\begin{center}
$ \pi_i = p_i*R_i + (1-p_i)* T_i - e_i$
\end{center}

We then assume that the returns to losing for the higher-ability contestant depend on $\theta_h$ ($\theta_h^{-\alpha} * T_h$). Thus, the higher-ability contestant receives a decreasing reward for losing with increasing contestant heterogeneity. This yields the following equilibrium effort:

\begin{center}
    $e_l^* = \frac{\theta_h^{\alpha+1} (R_l-T_l)^2 (\theta_h^\alpha R_h - T_h)}{(R_l\theta_h^\alpha - T_l\theta_h^\alpha + R_h\theta_h^{\alpha+1}-\theta_h T_h)^2} $ and $e_h^* = \frac{\theta_h (R_l-T_l) (\theta_h^\alpha R_h - T_h)^2}{(R_l\theta_h^\alpha - T_l\theta_h^\alpha + R_h\theta_h^{\alpha+1}-\theta_h T_h)^2} $.
\end{center}

We illustrate a numerical example assuming $\alpha = 0.2$ and $T_1 = T_2 = 0.5$ in Figure \ref{subfig:model5} and show that this model also aligns with the observed result in Section \ref{subsec:cont_het}.\footnote{Note that with $T_1 = T_2 = 0$, this model is equal to our baseline model.}

\subsection{Alternative explanations}\label{subsec:adisc}

An alternative potential mechanism could be that the higher-ability contestant tries to save resources when playing against a weaker opponent for subsequent rounds. Even though contests that include heterogeneous contestants are mechanically shorter than contests with homogeneous participants, the higher-ability contestant may try to "finish as early as possible" in heterogeneous contests to save resources and increase rest time until the next round. We think this is unlikely to be the case, since there is usually enough rest time between contests and because contests against much weaker opponents are already mechanically shorter without any increase in performance. However, we also empirically address this idea. First, to make "finishing as early as possible" a plausible strategy, we should observe a negative relationship between the length of previous contests and the current performance. In Table \ref{apptab:table_main_perf_sens}, we show that the number of legs played in previous rounds hardly correlates with the current match's performance. Second, if saving resources would be a motivation, we should observe this behavior, particularly before difficult contests (strong future opponents), because rest time is more relevant in such cases. In Table \ref{apptab:table_resource}, we show that the strength of the future opponent is, if anything, negatively correlated with our treatment effect. In sum, this suggests that "finishing early" to save resources is unlikely to explain our result.\par 

Finally, we consider risk strategies. In Section \ref{sec:data}, we have argued that there is little space for strategic adjustments in the first throws of a leg. In theory, minor strategic considerations are possible in the first throws in a leg. Highly skilled contestants typically aim for triple 20 to maximize points. In contrast, low-skilled contestants may prefer triple 19 because the neighboring fields (3 and 7) yield more points compared to the neighboring fields of the 20 (1 and 5) \cite{Tibshirani.2011}. Against a much weaker opponent, a skilled contestant might choose triple 19 to reduce variability and minimize the risk of losing due to bad luck. We think this is unlikely to explain our result for the following reasons. First, our contestants are all professional (highly skilled, low-variance) darts players, and the above argument applies only to high-variance players. Second, all contestants, most of the time, aim for triple 20 \cite{Klein.2020}. Third, anecdotal evidence also suggests that our contestants have "preferred" target fields that they routinely target and do not adjust this based on match circumstances. Fourth, three darts in the triple 20 (180) yield a particularly positive crowd response, which anecdotally darts players try to achieve and celebrate.

\section{Conclusion}\label{sec:conc}

Many situations can be characterized as tournament-style competitions. Tournament theory suggests that large inherent differences in the contestants' ability can adversely affect effort provision and performance. We use a rich panel data set of professional darts players to show that only contest designers with particular objective functions may fear the adverse effects of heterogeneity in the ability among contestants. \par 

We show that increasing heterogeneity among contestants has detrimental effects on the performance of lower-ability contestants. This is in line with the theoretical predictions. However, the contest designer interested in the performance of higher-ability contestants may find positive returns to increasing heterogeneity among contestants. This result is mainly driven by the highest performers in the ability distribution. In addition, we show that the adverse incentive effects on the lower-ability contestant can be mitigated by biasing the contest towards the lower-ability contestant, as done in many affirmative action policies. We also demonstrate substantial shadow effects across tournament stages. Higher-ability contestants notably underperform as the ability of future competitors increases, reducing their advancement chances. In contrast, lower-ability contestants remain unaffected by future opponents' strength.\par 

We offer a theory-driven discussion on possible mechanisms to explain the increased performance of the higher-ability contestant with increasing contestant heterogeneity. We demonstrate that out-of-equilibrium beliefs, differential reward evaluations, and risk and effort-saving strategies are unlikely to account for the result. Two potential explanations are choking under pressure \cite{Baumeister.1984} and anticipated regret \cite{Hyndman.2012}. Using a two-player Tullock contest model, we illustrate how the pressures of close competitions and the anticipation of regret from losing to a significantly weaker opponent can explain the observed performance patterns of the higher-ability contestant.\par

Our findings are relevant in various contexts, including organizations or sales departments that use internal competition as an incentive mechanism. In such cases, managers should be conscious that differences in contestants' abilities can lead to anticipated adverse effects on performance, particularly among employees with lower abilities. In terms of the performance of the lower-ability contestants, these managers may see some benefit in biasing the contest in favor of the lower-ability contestants. Conversely, principals interested in the performance of the higher-ability contestants, such as, for example, innovation tournament designers or public administrators planning public tenders, may view contestant heterogeneity differently as their primary objective is to encourage the highest level of performance \cite{Terwiesch.2009b}. In these cases, some heterogeneity in abilities among contestants may be seen as an asset rather than a liability. However, concerning the optimal design of such a contest, they should consider meaningful shadow effects across stages on the performance of the higher-ability contestant. \par

Given the prevalence of tournament scenarios, it is crucial to comprehend the impact of key contest design parameters on performance. Our findings contribute to this understanding by shedding light on how contestant heterogeneity influences performance in tournament-type situations. This insight is a valuable stride toward optimizing situations where individuals with varying abilities compete within relative reward frameworks.\par

\newpage

@article{Ammann.2016,
 author = {Ammann, Manuel and Horsch, Philipp and Oesch, David},
 title = {Competing with Superstars},
 journal = {Management Science},
 volume = {62},
 number = {10},
 pages = {2842-2858},
 year = {2016}
}

@article{Ales.2017,
 title = {Optimal award scheme in innovation tournaments},
 author = {Ales, Laurence and Cho, Soo-Haeng and K{\"o}rpeo{\u{g}}lu, Ersin},
 journal = {Operations Research},
 volume = {65},
 number = {3},
 pages = {693-702},
 year = {2017},
 publisher = {INFORMS}
}

@article{Anderson.2007,
author = {Anderson, Axel and Cabral, Luís M. B.},
title = {Go for broke or play it safe? Dynamic competition with choice of variance},
journal = {The RAND Journal of Economics},
volume = {38},
number = {3},
pages = {593-609},
year = {2007}
}

@article{Apesteguia.2010,
 Author = {Apesteguia, Jose and Palacios-Huerta, Ignacio},
 Title = {Psychological Pressure in Competitive Environments: Evidence from a Randomized Natural Experiment},
 Journal = {American Economic Review},
 Volume = {100},
 Number = {5},
 Year = {2010},
 Pages = {2548-2564}
}

@article{Ariely.2009,
 author = {Dan Ariely and Uri Gneezy and George Loewenstein and Nina Mazar},
 journal = {The Review of Economic Studies},
 number = {2},
 pages = {451-469},
 publisher = {[Oxford University Press, The Review of Economic Studies, Ltd.]},
 title = {Large Stakes and Big Mistakes},
 urldate = {2023-10-23},
 volume = {76},
 year = {2009}
}

@article{Bach.2009,
 author = {Norbert Bach and Oliver Gürtler and Joachim Prinz},
 journal = {Management Revue},
 number = {3},
 pages = {239-253},
 publisher = {Rainer Hampp Verlag},
 title = "{Incentive effects in tournaments with heterogeneous competitors — An analysis of the olympic rowing regatta in sydney 2000}",
 volume = {20},
 year = {2009}
}

@article{Baik.1994,
  title={Effort levels in contests with two asymmetric players},
  author={Baik, Kyung Hwan},
  journal={Southern Economic Journal},
  pages={367-378},
  volume = {61},
  number= {2},
  year={1994},
  publisher={JSTOR}
}

@article{Baumeister.1984,
  title={Choking under pressure: self-consciousness and paradoxical effects of incentives on skillful performance.},
  author={Baumeister, Roy F},
  journal={Journal of Personality and Social Psychology},
  volume={46},
  number={3},
  pages={610-620},
  year={1984},
  publisher={American Psychological Association}
}

@article{Baye.1993,
 author = {Michael R. Baye and Dan Kovenock and Casper G. de Vries},
 journal = {The American Economic Review},
 number = {1},
 pages = {289-294},
 publisher = {American Economic Association},
 title = {Rigging the Lobbying Process: An Application of the All-Pay Auction},
 urldate = {2023-10-31},
 volume = {83},
 year = {1993}
}

@article{Becker.1992,
  title={The incentive effects of tournament compensation systems},
  author={Becker, Brian E and Huselid, Mark A},
  journal={Administrative Science Quarterly},
  pages={336-350},
  year={1992},
  publisher={JSTOR}
}

@article{Hyndman.2012,
title = {Rent seeking with regretful agents: Theory and experiment},
journal = {Journal of Economic Behavior \& Organization},
volume = {84},
number = {3},
pages = {866-878},
year = {2012},
issn = {0167-2681},
author = {Kyle Hyndman and Erkut Y. Ozbay and Pacharasut Sujarittanonta}
}

@article{Bernard.2010,
title = {Level-k reasoning in contests},
journal = {Economics Letters},
volume = {108},
number = {2},
pages = {149-152},
year = {2010},
author = {Mark Bernard},
keywords = {Contests, Cognitive hierarchy},
abstract = {We introduce level-k reasoning to contest theory, compare its predictions to those of Nash Equilibrium and relate to the experimental evidence.}
}

@article{Berger.2016,
 author = {Johannes Berger and Petra Nieken},
 title ={Heterogeneous Contestants and the Intensity of Tournaments: An Empirical Investigation},
 journal = {Journal of Sports Economics},
 volume = {17},
 number = {7},
 pages = {631-660},
 year = {2016}
}

@article{Bilen.2023,
 title = {The Queen's Gambit: Explaining the superstar effect using evidence from chess},
 journal = {Journal of Economic Behavior \& Organization},
 volume = {215},
 pages = {307-324},
 year = {2023},
 issn = {0167-2681},
 author = {Eren Bilen and Alexander Matros},
 keywords = {Superstar, Tournament, Effort, Peer-effect, Chess}
}

@article{Bimpikis.2019,
author = {Bimpikis, Kostas and Ehsani, Shayan and Mostagir, Mohamed},
title = {Designing Dynamic Contests},
journal = {Operations Research},
volume = {67},
number = {2},
pages = {339-356},
year = {2019}
}

@article{Boudreau.2011,
  title={Incentives and problem uncertainty in innovation contests: An empirical analysis},
  author={Boudreau, Kevin J and Lacetera, Nicola and Lakhani, Karim R},
  journal={Management Science},
  volume={57},
  number={5},
  pages={843-863},
  year={2011},
  publisher={INFORMS}
}

@article{Boudreau.2016,
  title={Performance responses to competition across skill levels in rank-order tournaments: Field evidence and implications for tournament design},
  author={Boudreau, Kevin J and Lakhani, Karim R and Menietti, Michael},
  journal={The RAND Journal of Economics},
  volume={47},
  number={1},
  pages={140-165},
  year={2016},
  publisher={Wiley Online Library}
}

@article{Brown.2011,
 ISSN = {00223808, 1537534X},
 author = {Brown, Jennifer},
 journal = {Journal of Political Economy},
 number = {5},
 pages = {982-1013},
 publisher = {The University of Chicago Press},
 title = {Quitters Never Win: The (Adverse) Incentive Effects of Competing with Superstars},
 volume = {119},
 year = {2011}
}

@article{Brown.2014,
 author = {Brown, Jennifer and Minor, Dylan B.},
 journal = {Management Science},
 number = {12},
 pages = {3087-3102},
 publisher = {INFORMS},
 title = "{Selecting the best? Spillover and shadows in elimination tournaments}",
 volume = {60},
 year = {2014}
}

@article{Brown.2017,
title = {The hidden perils of affirmative action: Sabotage in handicap contests},
journal = {Journal of Economic Behavior \& Organization},
volume = {133},
pages = {273-284},
year = {2017},
author = {Brown, Alasdair and Chowdhury, Subhasish M.},
keywords = {Sabotage, Contests, Contest design, Superstars, Handicapping, Horse racing}
}

@article{Bull.1987,
 author = {Clive Bull and Andrew Schotter and Keith Weigelt},
 journal = {Journal of Political Economy},
 number = {1},
 pages = {1-33},
 publisher = {University of Chicago Press},
 title = {Tournaments and Piece Rates: An Experimental Study},
 volume = {95},
 year = {1987}
}

@article{Calsamiglia.2013,
 title = {The incentive effects of affirmative action in a real-effort tournament},
 journal = {Journal of Public Economics},
 volume = {98},
 pages = {15-31},
 year = {2013},
 issn = {0047-2727},
 author = {Caterina Calsamiglia and Jörg Franke and Pedro Rey-Biel},
 keywords = {Affirmative action, Tournament, Real-effort, Experiment, Sudoku}
}

@article{Camerer.2005,
  title={Neuroeconomics: How neuroscience can inform economics},
  author={Camerer, Colin and Loewenstein, George and Prelec, Drazen},
  journal={Journal of Economic Literature},
  volume={43},
  number={1},
  pages={9-64},
  year={2005},
  publisher={American Economic Association}
}

@article{Cason.2010,
 title = {Entry into winner-take-all and proportional-prize contests: An experimental study},
 journal = {Journal of Public Economics},
 volume = {94},
 number = {9-10},
 pages = {604-611},
 year = {2010},
 issn = {0047-2727},
 author = {Timothy N. Cason and William A. Masters and Roman M. Sheremeta},
 keywords = {Performance pay, Tournament, Piece rate, Tournament design, Contest, Experiments, Risk aversion, Feedback, Gender}
}

@article{CasasArce.2009,
 title={Relative performance compensation, contests, and dynamic incentives},
 author={Casas-Arce, Pablo and Martinez-Jerez, F Asis},
 journal={Management Science},
 volume={55},
 number={8},
 pages={1306-1320},
 year={2009},
 publisher={INFORMS}
}

@article{Chan.2014,
 author = {Chan, Tat Y. and Li, Jia and Pierce, Lamar},
 title = {Compensation and Peer Effects in Competing Sales Teams},
 journal = {Management Science},
 volume = {60},
 number = {8},
 pages = {1965-1984},
 year = {2014}
}

@article{Che.2003,
 Author = {Che, Yeon-Koo and Gale, Ian},
 Title = {Optimal Design of Research Contests },
 Journal = {American Economic Review},
 Volume = {93},
 Number = {3},
 Year = {2003},
 Pages = {646-671}
}

@article{Chowdhury.2023,
 author = {Chowdhury, Subhasish M. and Esteve-González, Patricia and Mukherjee, Anwesha},
 title = {Heterogeneity, leveling the playing field, and affirmative action in contests},
 journal = {Southern Economic Journal},
 volume = {89},
 number = {3},
 pages = {924-974},
 keywords = {affirmative action, contest, heterogeneity, survey},
 year = {2023}
}

@inbook {Coates.2018,
 author = "Dennis Coates and Brad R. Humphreys",
 title = "Behavioral and sports economics",
 booktitle = "Handbook of Behavioral Industrial Organization",
 year = "2018",
 publisher = "Edward Elgar Publishing",
 address = "Cheltenham, UK",
 isbn = "9781784718978"
}

@article{Connelly.2014,
  title={Tournament theory: Thirty years of contests and competitions},
  author={Connelly, Brian L and Tihanyi, Laszlo and Crook, T Russell and Gangloff, K Ashley},
  journal={Journal of Management},
  volume={40},
  number={1},
  pages={16-47},
  year={2014},
  publisher={Sage Publications Sage CA: Los Angeles, CA}
}

@incollection{Corchon.2018,
  title={Contest theory},
  author={Corch{\'o}n, Luis C and Serena, Marco},
  booktitle="{Handbook of Game Theory and Industrial Organization, Volume II}",
  year={2018},
  publisher={Edward Elgar Publishing}
}

@article{Dechenaux.2015,
  title={A survey of experimental research on contests, all-pay auctions and tournaments},
  author={Dechenaux, Emmanuel and Kovenock, Dan and Sheremeta, Roman M},
  journal={Experimental Economics},
  volume={18},
  pages={609-669},
  year={2015},
  publisher={Springer}
}

@article{Deutscher.2023,
  title={Who’s afraid of the GOATs?-Shadow Effects of Tennis Superstars},
  author={Deutscher, Christian and Neuberg, Lena and Thiem, Stefan},
  journal={Journal of Economic Psychology},
  pages={102663},
  year={2023},
  publisher={Elsevier}
}

@article{Dohmen.2008,
  title={Do professionals choke under pressure?},
  author={Dohmen, Thomas J},
  journal={Journal of Economic Behavior \& Organization},
  volume={65},
  number={3-4},
  pages={636-653},
  year={2008},
  publisher={Elsevier}
}

@article{Drugov.2017,
title = {Biased contests for symmetric players},
journal = {Games and Economic Behavior},
volume = {103},
pages = {116-144},
year = {2017},
issn = {0899-8256},
author = {Mikhail Drugov and Dmitry Ryvkin},
keywords = {Biased contest, Biased contest success function, Aggregate effort, Predictive power, Winner's effort}
}

@article{Drugov.2020,
title = {Tournament rewards and heavy tails},
journal = {Journal of Economic Theory},
volume = {190},
pages = {105116},
year = {2020},
issn = {0022-0531},
author = {Mikhail Drugov and Dmitry Ryvkin},
keywords = {Heavy tails, Power law, Tournament, Optimal allocation of prizes, Failure rate}
}

@article{Drugov.2022,
  title={Hunting for the discouragement effect in contests},
  author={Drugov, Mikhail and Ryvkin, Dmitry},
  journal={Review of Economic Design},
  pages={1-27},
  year={2022},
  publisher={Springer}
}

@article{Ehrenberg.1990a,
  title={Do tournaments have incentive effects?},
  author={Ehrenberg, Ronald G and Bognanno, Michael L},
  journal={Journal of Political Economy},
  volume={98},
  number={6},
  pages={1307-1324},
  year={1990},
  publisher={The University of Chicago Press}
}

@article{Ehrenberg1990b,
  title={The incentive effects of tournaments revisited: Evidence from the European PGA tour},
  author={Ehrenberg, Ronald G and Bognanno, Michael L},
  journal={ILR Review},
  volume={43},
  number={3},
  pages={74-88},
  year={1990},
  publisher={SAGE Publications Sage CA: Los Angeles, CA}
}

@article{Ewerhart.2017,
  title={Revenue ranking of optimally biased contests: The case of two players},
  author={Ewerhart, Christian},
  journal={Economics Letters},
  volume={157},
  pages={167-170},
  year={2017},
  publisher={Elsevier}
}

@article{Fang.2020,
author = {Fang, Dawei and Noe, Thomas and Strack, Philipp},
title = {Turning up the Heat: The Discouraging Effect of Competition in Contests},
journal = {Journal of Political Economy},
volume = {128},
number = {5},
pages = {1940-1975},
year = {2020},
}

@article{Fehr.2018,
  title={Exclusion in all-pay auctions: An experimental investigation},
  author={Fehr, Dietmar and Schmid, Julia},
  journal={Journal of Economics \& Management Strategy},
  volume={27},
  number={2},
  pages={326-339},
  year={2018},
  publisher={Wiley Online Library}
}

@article{Fershtman.2011,
 ISSN = {15424766, 15424774},
 URL = {http://www.jstor.org/stable/25836068},
 author = {Chaim Fershtman and Uri Gneezy},
 journal = {Journal of the European Economic Association},
 number = {2},
 pages = {318-336},
 publisher = {Oxford University Press},
 title = {THE TRADEOFF BETWEEN PERFORMANCE AND QUITTING IN HIGH POWER TOURNAMENTS},
 urldate = {2023-05-02},
 volume = {9},
 year = {2011}
}

@article{Filiz.2007,
Author = {Filiz-Ozbay, Emel and Ozbay, Erkut Y.},
Title = {Auctions with Anticipated Regret: Theory and Experiment},
Journal = {American Economic Review},
Volume = {97},
Number = {4},
Year = {2007},
Pages = {1407–1418}}

@article{filiz.2010,
title = {Anticipated loser regret in third price auctions},
journal = {Economics Letters},
volume = {107},
number = {2},
pages = {217-219},
year = {2010},
author = {Emel Filiz-Ozbay and Erkut Y. Ozbay},
keywords = {Overbidding, Third price auction, Regret}}
}

@article{Franke.2012,
  title={The incentive effects of levelling the playing field--an empirical analysis of amateur golf tournaments},
  author={Franke, J{\"o}rg},
  journal={Applied Economics},
  volume={44},
  number={9},
  pages={1193-1200},
  year={2012},
  publisher={Taylor \& Francis}
}

@article{Franke.2013,
  title={Effort maximization in asymmetric contest games with heterogeneous contestants},
  author={Franke, J{\"o}rg and Kanzow, Christian and Leininger, Wolfgang and Schwartz, Alexandra},
  journal={Economic Theory},
  volume={52},
  pages={589-630},
  year={2013},
  publisher={Springer}
}

@article{Franke.2018,
 title = {Optimal favoritism in all-pay auctions and lottery contests},
 journal = {European Economic Review},
 volume = {104},
 pages = {22-37},
 year = {2018},
issn = {0014-2921},
 author = {Jörg Franke and Wolfgang Leininger and Cédric Wasser},
 keywords = {All-pay auction, Lottery contest, Favoritism, Head start, Revenue dominance}
}

@incollection{Fu.2019,
  title={Contests: Theory and topics},
  author={Fu, Qiang and Wu, Zenan},
  booktitle={Oxford Research Encyclopedia of Economics and Finance},
  year={2019}
}

@article{Fu.2012,
  title={The optimal multi-stage contest},
  author={Fu, Qiang and Lu, Jingfeng},
  journal={Economic Theory},
  volume={51},
  pages={351-382},
  year={2012},
  publisher={Springer}
}

@article{Fu.2020,
  title={On the optimal design of biased contests},
  author={Fu, Qiang and Wu, Zenan},
  journal={Theoretical Economics},
  volume={15},
  number={4},
  pages={1435-1470},
  year={2020},
  publisher={Wiley Online Library}
}

@article{Fullerton.1999,
 ISSN = {00223808, 1537534X},
 author = {Richard L. Fullerton and R. Preston McAfee},
 journal = {Journal of Political Economy},
 number = {3},
 pages = {573-605},
 publisher = {The University of Chicago Press},
 title = {Auctionin Entry into Tournaments},
 urldate = {2023-04-25},
 volume = {107},
 year = {1999}
}

@article{Gauriot.2019,
    author = {Gauriot, Romain and Page, Lionel},
    title = "{Does success breed success? A quasi-experiment on strategic momentum in dynamic contests}",
    journal = {The Economic Journal},
    volume = {129},
    number = {624},
    pages = {3107-3136},
    year = {2019},
    ISSN = {0013-0133}
}

@article{Genakos.2012,
author = {Genakos, Christos and Pagliero, Mario},
title = {Interim Rank, Risk Taking, and Performance in Dynamic Tournaments},
journal = {Journal of Political Economy},
volume = {120},
number = {4},
pages = {782-813},
year = {2012}
}

@article{Gilsdorf.2008,
  title={Tournament incentives and match outcomes in women's professional tennis},
  author={F. Gilsdorf, Keith and Sukhatme, Vasant A},
  journal={Applied Economics},
  volume={40},
  number={18},
  pages={2405-2412},
  year={2008},
  publisher={Taylor \& Francis}
}

@article{Gill.2012,
  title={A structural analysis of disappointment aversion in a real effort competition},
  author={Gill, David and Prowse, Victoria},
  journal={American Economic Review},
  volume={102},
  number={1},
  pages={469-503},
  year={2012},
  publisher={American Economic Association}
}

@article{Goller.2023,
author = {Goller, Daniel},
journal = {Annals of Operations Research},
title = {{Analysing a built-in advantage in asymmetric darts contests using causal machine learning}},
volume = {325},
number = {1},
pages = {649-679},
year = {2023}
}

@article{Goriaev.2001,
  title={Mutual fund tournament: Risk taking incentives induced by ranking objectives},
  author={Goriaev, Alexei and Palomino, Frederic and Prat, Andrea},
  journal={Available at SSRN 270304},
  year={2001}
}

@article{Gross.2020,
  title={Creativity under fire: The effects of competition on creative production},
  author={Gross, Daniel P},
  journal={Review of Economics and Statistics},
  volume={102},
  number={3},
  pages={583-599},
  year={2020},
  publisher={MIT Press One Rogers Street, Cambridge, MA 02142-1209, USA journals-info~…}
}

@article{Gurtler.2015,
author = {G\"{u}rtler, Marc and G\"{u}rtler, Oliver},
title = {The Optimality of Heterogeneous Tournaments},
journal = {Journal of Labor Economics},
volume = {33},
number = {4},
pages = {1007-1042},
year = {2015}
}

@article{Grund.2005,
author = {Christian Grund and Oliver Gürtler},
title = {An empirical study on risk-taking in tournaments},
journal = {Applied Economics Letters},
volume = {12},
number = {8},
pages = {457--461},
year = {2005},
publisher = {Routledge}
}

@article{Harb.2019,
title = {Choking under pressure in front of a supportive audience: Evidence from professional biathlon},
journal = {Journal of Economic Behavior \& Organization},
volume = {166},
pages = {246-262},
year = {2019},
author = {Ken Harb-Wu and Alex Krumer},
keywords = {Choking under pressure, Paradoxical performance effects of incentives, Social pressure, Biathlon, Home advantage}
}

@article{Harbaugh.2005,
  title={Early round upsets and championship blowouts},
  author={Harbaugh, Rick and Klumpp, Tilman},
  journal={Economic Inquiry},
  volume={43},
  number={2},
  pages={316-329},
  year={2005},
  publisher={Wiley Online Library}
}

@article{Harbring.2003,
  title={An experimental study on tournament design},
  author={Harbring, Christine and Irlenbusch, Bernd},
  journal={Labour Economics},
  volume={10},
  number={4},
  pages={443-464},
  year={2003},
  publisher={Elsevier}
}

@article{Harris.1987,
 ISSN = {00346527, 1467937X},
 author = {Christopher Harris and John Vickers},
 journal = {The Review of Economic Studies},
 number = {1},
 pages = {1-21},
 publisher = {[Oxford University Press, Review of Economic Studies, Ltd.]},
 title = {Racing with Uncertainty},
 urldate = {2023-10-31},
 volume = {54},
 year = {1987}
}

@article{Hill.2014,
  title={The superstar effect in 100-meter tournaments},
  author={Hill, Brian},
  journal={International Journal of Sport Finance},
  volume={9},
  number={2},
  pages={111},
  year={2014},
  publisher={Fitness Information Technology, A Division of ICPE West Virginia University}
}

@article{Hill.2018,
  title={Shadow and spillover effects of competition in NBA playoffs},
  author={Hill, Brian},
  journal={Journal of Sports Economics},
  volume={19},
  number={8},
  pages={1067-1092},
  year={2018},
  publisher={Sage Publications Sage CA: Los Angeles, CA}
}

@article{Hörner.2004,
 ISSN = {00346527, 1467937X},
 URL = {http://www.jstor.org/stable/3700728},
 author = {Johannes Hörner},
 journal = {The Review of Economic Studies},
 number = {4},
 pages = {1065-1088},
 publisher = {[Oxford University Press, Review of Economic Studies, Ltd.]},
 title = {A Perpetual Race to Stay Ahead},
 urldate = {2023-05-02},
 volume = {71},
 year = {2004}
}

@article{kirkegaard.2012,
  title={Favoritism in asymmetric contests: Head starts and handicaps},
  author={Kirkegaard, Ren{\'e}},
  journal={Games and Economic Behavior},
  volume={76},
  number={1},
  pages={226-248},
  year={2012},
  publisher={Elsevier}
}

@article{kirkegaard.2023,
  title={Contest design with stochastic performance},
  author={Kirkegaard, Ren{\'e}},
  journal={American Economic Journal: Microeconomics},
  volume={15},
  number={1},
  pages={201-238},
  year={2023},
  publisher={American Economic Association 2014 Broadway, Suite 305, Nashville, TN 37203-2425}
}

@article{Kleinknecht.2021,
author = {Kleinknecht, Janina and Ulmer, Clara},
title = {Under the shadow of the future: Gender-specific reactions to (un)certain future interactions},
journal = {Journal of Economics \& Management Strategy},
year = {2023}
}

@article{Klein.2020,
title = {Incentives, performance and choking in darts},
journal = {Journal of Economic Behavior \& Organization},
volume = {169},
pages = {38-52},
year = {2020},
author = {Bouke {Klein Teeselink} and Rogier J.D. {Potter van Loon} and Martijn J. {van den Assem} and Dennie {van Dolder}},
}

@article{Kimbrough.2014,
title = {When parity promotes peace: Resolving conflict between asymmetric agents},
journal = {Journal of Economic Behavior \& Organization},
volume = {99},
pages = {96-108},
year = {2014},
ISSN = {0167-2681},
author = {Erik O. Kimbrough and Roman M. Sheremeta and Timothy W. Shields},
keywords = {Contest, Asymmetries, Conflict resolution, Experiments}
}

@article{Dato.2018,
  title={Expectation-based loss aversion and rank-order tournaments},
  author={Dato, Simon and Grunewald, Andreas and M{\"u}ller, Daniel},
  journal={Economic Theory},
  volume={66},
  pages={901-928},
  year={2018},
  publisher={Springer}
}

@article{Konrad.2009,
title = {Multi-battle contests},
journal = {Games and Economic Behavior},
volume = {66},
number = {1},
pages = {256-274},
year = {2009},
ISSN = {0899-8256},
author = {Kai A. Konrad and Dan Kovenock},
keywords = {All-pay auction, Contest, Race, Conflict, Multi-stage, R&D, Endogenous uncertainty, Preemption, Discouragement}
}

@article{Konrad.2009a,
  title={Strategy and dynamics in contests},
  author={Konrad, Kai A.},
  journal={OUP Catalogue},
  year={2009},
  pages={No. 9780199549603},
  publisher={Oxford University Press}
}

@article{Kubitz.2023,
Author = {Kubitz, Greg},
Title = {Two-Stage Contests with Private Information},
Journal = {American Economic Journal: Microeconomics},
Volume = {15},
Number = {1},
Year = {2023},
Pages = {239-287}
}

@article{Lackner.2020,
title = "{Are competitors forward looking in strategic interactions? Field evidence from multistage tournaments}",
journal = {Journal of Economic Behavior \& Organization},
volume = {179},
pages = {544-565},
year = {2020},
issn = {0167-2681},
author = {Mario Lackner and Rudi Stracke and Uwe Sunde and Rudolf Winter-Ebmer},
keywords = {Promotion tournament, Multistage contest, Elimination, Forward-looking behavior, Heterogeneity},
}

@article{Lazear.1981,
 author = {Edward P. Lazear and Sherwin Rosen},
 journal = {Journal of Political Economy},
 number = {5},
 pages = {841-864},
 publisher = {University of Chicago Press},
 title = {Rank-Order Tournaments as Optimum Labor Contracts},
 urldate = {2022-04-13},
 volume = {89},
 year = {1981}
}

@article{Lazear.2018,
Author = {Lazear, Edward P.},
Title = {Compensation and Incentives in the Workplace},
Journal = {Journal of Economic Perspectives},
Volume = {32},
Number = {3},
Year = {2018},
Pages = {195-214}
}

@article{Leibbrandt.2018,
author = {Leibbrandt, Andreas and Wang, Liang Choon and Foo, Cordelia},
title = {Gender Quotas, Competitions, and Peer Review: Experimental Evidence on the Backlash Against Women},
journal = {Management Science},
volume = {64},
number = {8},
pages = {3501-3516},
year = {2018}
}

@article{Letina.2023,
  title={Optimal contest design: Tuning the heat},
  author={Letina, Igor and Liu, Shuo and Netzer, Nick},
  journal={Journal of Economic Theory},
  pages={105616},
  year={2023},
  publisher={Elsevier}
}

@article{Lynch.2000,
  title={The rewards to running: Prize structure and performance in professional road racing},
  author={Lynch, James G and Zax, Jeffrey S},
  journal={Journal of Sports Economics},
  volume={1},
  number={4},
  pages={323-340},
  year={2000},
  publisher={Sage Publications Sage CA: Thousand Oaks, CA}
}

@article{Lynch.2005,
  title={The effort effects of prizes in the second half of tournaments},
  author={Lynch, James G},
  journal={Journal of Economic Behavior \& Organization},
  volume={57},
  number={1},
  pages={115-129},
  year={2005},
  publisher={Elsevier}
}

@article{Knoeber.1994,
  title={Testing the theory of tournaments: An empirical analysis of broiler production},
  author={Knoeber, Charles R and Thurman, Walter N},
  journal={Journal of labor economics},
  volume={12},
  number={2},
  pages={155-179},
  year={1994},
  publisher={University of Chicago Press}
}

@article{Kraekel.2004,
  title={Risk taking in asymmetric tournaments},
  author={Kr{\"a}kel, Matthias and Sliwka, Dirk},
  journal={German Economic Review},
  volume={5},
  number={1},
  pages={103-116},
  year={2004},
  publisher={Wiley Online Library}
}

@article{Mago.2019,
title = {Best-of-five contest: An experiment on gender differences},
journal = {Journal of Economic Behavior \& Organization},
volume = {162},
pages = {164-187},
year = {2019},
issn = {0167-2681},
doi = {https://doi.org/10.1016/j.jebo.2019.04.015},
url = {https://www.sciencedirect.com/science/article/pii/S0167268119301179},
author = {Shakun D. Mago and Laura Razzolini},
keywords = {Lottery contest, Laboratory experiment, Gender differences, Confidence}
}

@article{Malueg.2010,
 author = {David A. Malueg and Andrew J. Yates},
 journal = {The Review of Economics and Statistics},
 number = {3},
 pages = {689-692},
 publisher = {The MIT Press},
 title = {TESTING CONTEST THEORY: EVIDENCE FROM BEST-OF-THREE TENNIS MATCHES},
 urldate = {2023-10-31},
 volume = {92},
 year = {2010}
}

@article{Morgan.2022,
  title={The limits of meritocracy},
  author={Morgan, John and Tumlinson, Justin and V{\'a}rdy, Felix},
  journal={Journal of Economic Theory},
  volume={201},
  pages={105414},
  year={2022},
  publisher={Elsevier}
}

@article{Moldovanu.2001,
 author = {Benny Moldovanu and Aner Sela},
 journal = {The American Economic Review},
 number = {3},
 pages = {542-558},
 publisher = {American Economic Association},
 title = {The Optimal Allocation of Prizes in Contests},
 urldate = {2023-05-02},
 volume = {91},
 year = {2001}
}

@article{Muller.2010,
 author = {Wieland Müller and Andrew Schotter},
 journal = {Journal of the European Economic Association},
 number = {4},
 pages = {717--743},
 publisher = {Oxford University Press},
 title = {Workaholics and dropouts in organizations},
 volume = {8},
 year = {2010}
}

@article{Nalebuff.1983,
 ISSN = {0361915X},
 URL = {http://www.jstor.org/stable/3003535},
 author = {Barry J. Nalebuff and Joseph E. Stiglitz},
 journal = {The Bell Journal of Economics},
 number = {1},
 pages = {21-43},
 publisher = {[RAND Corporation, Wiley]},
 title = {Prizes and Incentives: Towards a General Theory of Compensation and Competition},
 urldate = {2023-09-12},
 volume = {14},
 year = {1983}
}

@article{Neugart.2013,
title = {Sequential teamwork in competitive environments: Theory and evidence from swimming data},
journal = {European Economic Review},
volume = {63},
pages = {186-205},
year = {2013},
ISSN = {0014-2921},
doi = {https://doi.org/10.1016/j.euroecorev.2013.07.006},
url = {https://www.sciencedirect.com/science/article/pii/S0014292113000925},
author = {Michael Neugart and Matteo G. Richiardi},
keywords = {Team production, Contest, Intergroup competition, Sequential contribution, Free-riding}
}

@article{Oettinger.2020,
author = {Ötting, Marius and Langrock, Roland and Deutscher, Christian and Leos-Barajas, Vianey},
title = {The hot hand in professional darts},
journal = {Journal of the Royal Statistical Society: Series A (Statistics in Society)},
volume = {183},
number = {2},
pages = {565-580},
keywords = {Hidden Markov model, Hot hand, Sports statistics, State space model, Time series},
year = {2020}
}

@article{Pope.2011,
Author = {Pope, Devin G. and Schweitzer, Maurice E.},
Title = {Is Tiger Woods Loss Averse? Persistent Bias in the Face of Experience, Competition, and High Stakes},
Journal = {American Economic Review},
Volume = {101},
Number = {1},
Year = {2011},
Pages = {129-157}
}

@article{Prendergast.1999,
 ISSN = {00220515},
 URL = {http://www.jstor.org/stable/2564725},
 author = {Canice Prendergast},
 journal = {Journal of Economic Literature},
 number = {1},
 pages = {7-63},
 publisher = {American Economic Association},
 title = {The Provision of Incentives in Firms},
 urldate = {2022-04-13},
 volume = {37},
 year = {1999}
}

@article{Ridlon.2013,
author = {Ridlon, Robert and Shin, Jiwoong},
title = {Favoring the Winner or Loser in Repeated Contests},
journal = {Marketing Science},
volume = {32},
number = {5},
pages = {768-785},
year = {2013},
doi = {10.1287/mksc.2013.0798},
URL = {https://doi.org/10.1287/mksc.2013.0798},
eprint = {https://doi.org/10.1287/mksc.2013.0798},
}

@article{Rosen.1981,
 author = {Sherwin Rosen},
 journal = {The American Economic Review},
 number = {5},
 pages = {845-858},
 publisher = {American Economic Association},
 title = {The Economics of Superstars},
 urldate = {2022-04-13},
 volume = {71},
 year = {1981}
}

@article{Rosen.1986,
 abstract = {Contestants who succeed in attaining high ranks in elimination career ladders rest on their laurels in attempting to climb higher, unless top-ranking prizes are given a disproportionate weight in the purse. A large first-place prize gives survivors something to shoot for, independent of past performances and accomplishments.},
 author = {Sherwin Rosen},
 journal = {American Economic Review},
 number = {4},
 pages = {701-715},
 publisher = {American Economic Association},
 title = {Prizes and Incentives in Elimination Tournaments},
 volume = {76},
 year = {1986}
}

@article{Ryvkin.2008,
  title={The predictive power of three prominent tournament formats},
  author={Ryvkin, Dmitry and Ortmann, Andreas},
  journal={Management Science},
  volume={54},
  number={3},
  pages={492-504},
  year={2008},
  publisher={INFORMS}
}

@article{Ryvkin.2020,
author = {Ryvkin, Dmitry and Drugov, Mikhail},
title = {The shape of luck and competition in winner-take-all tournaments},
journal = {Theoretical Economics},
volume = {15},
number = {4},
pages = {1587-1626},
keywords = {Tournament, competition, heavy tails, stochastic number of players, unimodality, log supermodularity, failure rate, C72, D72, D82},
year = {2020}
}

@article{Ryvkin.2022,
author = {Ryvkin, Dmitry},
title = {To Fight or to Give Up? Dynamic Contests with a Deadline},
journal = {Management Science},
volume = {68},
number = {11},
pages = {8144-8165},
year = {2022},
doi = {10.1287/mnsc.2021.4206}
}

@article{Sahm.2022,
  title={Optimal accuracy of unbiased Tullock contests with two heterogeneous players},
  author={Sahm, Marco},
  journal={Games},
  volume={13},
  number={2},
  pages={24},
  year={2022},
  publisher={MDPI}
}

@article{Saguer.2023,
title = {Designing contests between heterogeneous contestants: An experimental study of tie-breaks and bid-caps in all-pay auctions},
journal = {European Economic Review},
volume = {154},
pages = {104327},
year = {2023},
ISSN = {0014-2921},
author = {Aniol Llorente-Saguer and Roman M. Sheremeta and Nora Szech},
keywords = {All-pay auction, Rent‑seeking, Bid-caps, Tie-breaks, Contest design}
}

@article{Schildberg.2023,
    author = {Schildberg-Hörisch, Hannah and Schwarz, Marco A and Trieu, Chi and Willrodt, Jana},
    title = "{Perceived fairness and consequences of affirmative action policies}",
    journal = {The Economic Journal},
    volume = {133},
    number = {656},
    pages = {3099-3135},
    year = {2023},   
    ISSN = {0013-0133}
}

@article{Schlosser.2019,
    author = {Schlosser, Analia and Neeman, Zvika and Attali, Yigal},
    title = "{Differential performance in high versus low stakes tests: Evidence from the gre test}",
    journal = {The Economic Journal},
    volume = {129},
    number = {623},
    pages = {2916-2948},
    year = {2019},
    ISSN = {0013-0133}
}

@article{Scholten.2023,
author = {Hendrik Scholten},
title ={You’ve Got Three Choices: Give in, Give up, or Give it All You’ve Got: Does Contest Heterogeneity Affect Effort in Individual Competitions?},
journal = {Journal of Sports Economics},
volume = {24},
number = {7},
pages = {932-965},
year = {2023},
abstract = {This study focuses on the effect of contest heterogeneity on individuals’ effort provision in rank-order tournaments. We add to the literature by analyzing data from a well-suited setting (professional darts tournaments) that differs significantly from those previously examined (mental effort task without direct interaction between contestants), while applying two different temporal types of heterogeneity, pre-contest and within-contest heterogeneity. We do not find consistent evidence for heterogeneity to influence players’ effort. }
}

@article{Schotter.1992,
 author = {Andrew Schotter and Keith Weigelt},
 journal = {The Quarterly Journal of Economics},
 number = {2},
 pages = {511-539},
 publisher = {Oxford University Press},
 title = {Asymmetric Tournaments, Equal Opportunity Laws, and Affirmative Action: Some Experimental Results},
 urldate = {2023-05-02},
 volume = {107},
 year = {1992}
}

@article{Muller.2010,
 ISSN = {15424766, 15424774},
 author = {Wieland Müller and Andrew Schotter},
 journal = {Journal of the European Economic Association},
 number = {4},
 pages = {717-743},
 publisher = {Oxford University Press},
 title = {WORKAHOLICS AND DROPOUTS IN ORGANIZATIONS},
 urldate = {2023-10-31},
 volume = {8},
 year = {2010}
}

@article{Nagel.1995,
  title={Unraveling in guessing games: An experimental study},
  author={Nagel, Rosemarie},
  journal={American Economic Review},
  volume={85},
  number={5},
  pages={1313-1326},
  year={1995},
  publisher={JSTOR}
}

@article{Segev2014,
  title={Multi-stage sequential all-pay auctions},
  author={Segev, Ella and Sela, Aner},
  journal={European Economic Review},
  volume={70},
  pages={371-382},
  year={2014},
  publisher={Elsevier}
}

@article{Sheremeta2010,
  title={Experimental comparison of multi-stage and one-stage contests},
  author={Sheremeta, Roman M},
  journal={Games and Economic Behavior},
  volume={68},
  number={2},
  pages={731-747},
  year={2010},
  publisher={Elsevier}
}

@article{Smith.2013,
 ISSN = {0022166X},
 author = {Jonathan Smith},
 journal = {The Journal of Human Resources},
 number = {2},
 pages = {265-285},
 publisher = {[University of Wisconsin Press, Board of Regents of the University of Wisconsin System]},
 title = {Peers, Pressure, and Performance at the National Spelling Bee},
 urldate = {2023-10-31},
 volume = {48},
 year = {2013}
}

@article{Stein.2002,
 ISSN = {00485829, 15737101},
 author = {William E. Stein},
 journal = {Public Choice},
 number = {3/4},
 pages = {325-336},
 publisher = {Springer},
 title = {Asymmetric Rent-Seeking with More than Two Contestants},
 urldate = {2023-04-13},
 volume = {113},
 year = {2002}
}

@article{Steinmayr.2018,
  title={Having the Lead vs Lagging Behind: The Incentive Effect of Handicaps in Tournaments},
  author={Steinmayr, Andreas and Stracke, Rudi and Zegners, Dainis},
  journal={Working paper},
  year={2018},
  publisher={SSRN}
}

@article{StPierre.2016,
  title={The role of inequality on effort in tournaments},
  author={St-Pierre, Marc},
  journal={Mathematical Social Sciences},
  volume={81},
  pages={38-52},
  year={2016},
  publisher={Elsevier}
}

@article{Sunde.2009,
  title={Heterogeneity and performance in tournaments: A test for incentive effects using professional tennis data},
  author={Sunde, Uwe},
  journal={Applied Economics},
  volume={41},
  number={25},
  pages={3199-3208},
  year={2009},
  publisher={Taylor \& Francis}
}

@article{Szymanski.2005,
title = {Incentive effects of second prizes},
journal = {European Journal of Political Economy},
volume = {21},
number = {2},
pages = {467-481},
year = {2005},
issn = {0176-2680},
author = {Stefan Szymanski and Tommaso M. Valletti},
keywords = {Imperfectly discriminating contests, Prizes, Logit contests, Rent-seeking},
}

@article{Tanaka.2012,
  title={Testing the incentive effects in tournaments with a superstar},
  author={Tanaka, Ryuichi and Ishino, Kazutoshi},
  journal={Journal of the Japanese and International Economies},
  volume={26},
  number={3},
  pages={393-404},
  year={2012},
  publisher={Elsevier}
}

@article{Terwiesch.2008,
author = {Terwiesch, Christian and Xu, Yi},
title = {Innovation Contests, Open Innovation, and Multiagent Problem Solving},
journal = {Management Science},
volume = {54},
number = {9},
pages = {1529-1543},
year = {2008}
}

@book{Terwiesch.2009b,
  title={Innovation tournaments: Creating and selecting exceptional opportunities},
  author={Terwiesch, Christian and Ulrich, Karl T},
  year={2009},
  publisher={Harvard Business Press}
}

@article{Tibshirani.2011,
  title={A statistician plays darts},
  author={Tibshirani, Ryan J and Price, Andrew and Taylor, Jonathan},
  journal={Journal of the Royal Statistical Society Series A: Statistics in Society},
  volume={174},
  number={1},
  pages={213-226},
  year={2011},
  publisher={Oxford University Press}
}

@article{Tullock.2001,
  title={Efficient rent seeking},
  author={Tullock, Gordon},
  journal={Efficient rent-seeking: Chronicle of an intellectual quagmire},
  pages={3-16},
  year={2001},
  publisher={Springer}
}

@article{Xiao.2016,
title = {Asymmetric all-pay contests with heterogeneous prizes},
journal = {Journal of Economic Theory},
volume = {163},
pages = {178-221},
year = {2016},
issn = {0022-0531},
author = {Jun Xiao},
keywords = {All-pay, Asymmetric, Contest, Heterogeneous}
}

@article{Yamane.2015,
 ISSN = {03470520, 14679442},
 author = {Shoko Yamane and Ryohei Hayashi},
 journal = {The Scandinavian Journal of Economics},
 number = {4},
 pages = {1230-1255},
 publisher = {[Wiley, The Scandinavian Journal of Economics]},
 title = {Peer Effects among Swimmers},
 urldate = {2023-10-14},
 volume = {117},
 year = {2015}
}

@article{Yerkes.1908,
author = {Yerkes, Robert M. and Dodson, John D.},
title = {The relation of strength of stimulus to rapidity of habit-formation},
journal = {Journal of Comparative Neurology and Psychology},
volume = {18},
number = {5},
pages = {459-482},
year = {1908}
}

@article{Breiman.2001,
  title={Random forests},
  author={Breiman, Leo},
  journal={Machine Learning},
  volume={45},
  number={1},
  pages={5-32},
  year={2001},
  publisher={Springer},
}

@article{Kennedy.2017,
  title={Non-parametric methods for doubly robust estimation of continuous treatment effects},
  author={Kennedy, Edward H and Ma, Zongming and McHugh, Matthew D and Small, Dylan S},
  journal={Journal of the Royal Statistical Society: Series B (Statistical Methodology)},
  volume={79},
  number={4},
  pages={1229-1245},
  year={2017},
  publisher={Wiley Online Library}
}

\vfill
\bibliographystyle{apacite}
\bibliography{references}

\newpage
\appendix
\beginappa
\section{Appendix}\label{sec:app}

\subsection{Descriptive Statistics}\label{subsec:app_desc}

	\begin{table}[H]
		\centering
		\caption{Full descriptive statistics} 
		\label{tab:table_app_desc}
		\begin{tabular}{l r r r} 
			\toprule \toprule

            \multicolumn{4}{l}{\textit{Panel A: Outcomes}} \vspace{0.7em}            \\ 
			Favorite wins                                & ~~~~ &  0.665 &           \\
		    Favorite 3-darts average (performance)       & &  102.254    & (9.057)   \\
		    Favorite 3-darts average in first half       & &  101.687    & (11.089)  \\
		    Favorite 3-darts average in second half      & &  102.892    & (11.454)  \\
		    Underdog 3-darts average (performance)       & &  97.595     & (8.879)   \\
		    Underdog 3-darts average in first half       & &  97.213     & (10.891)  \\
		    Underdog 3-darts average in second half      & &  98.016     & (11.369)  \\ 
		    Mean 3-darts average (performance)           & &  99.924     & (6.859)   \\ 
		    Mean 3-darts average in first half           & &  99.450     & (8.273)   \\ 
      		Mean 3-darts average in second half          & &  100.454    & (8.477)   \\
            Contest length (proportion of legs needed)   & &  0.790      & (0.159)   \\
            Number of max. performance (Score = 180) in contest & & 5.921 & (4.186)  \\ 
            \vspace{-0.5em}  \\
            \multicolumn{4}{l}{\textit{Panel B: Treatments}} \vspace{0.7em} \\ 
			Ability ratio (favorite to underdog)             & &   1.055     & (0.049)   \\
			Log ability difference (favorite minus underdog) & &   0.053     & (0.045)   \\
            Odds ability ratio (underdog to favorite)        & &   2.766     &  (2.387)  \\
            Expected ability (of opponent in next stage)     & &  94.695     & (3.963)   \\
   		    Underdog starts contest                          & &  0.481      &           \\
            \vspace{-0.5em}  \\
            \multicolumn{4}{l}{\textit{Panel C: Covariates}} \vspace{0.7em} \\ 
			Favorite ability                             & &  94.624     & (3.867)   \\
			Underdog ability                             & &  89.791     & (4.600)   \\
            Opponent is known                            & & 0.588       &           \\
   			Favorite starts contest                        & &  0.519      &           \\
      		Favorite world ranking                       & &  0.017      & (0.081) \\
         	Underdog world ranking                       & &  0.046      & (0.129) \\
            Favorite years playing darts (experience)    & &  21.333     & (8.921)   \\
            Underdog years playing darts (experience)    & &  19.766     & (10.493)  \\
            Favorite performs at home event              & &   0.034     &           \\
            Underdog performs at home event              & &   0.041     &           \\
            Prize money (standardized)                   & &   0.217     & (0.330)   \\
			\bottomrule \bottomrule
			\multicolumn{4}{l}{\footnotesize Notes: Mean and standard deviations (in parentheses; for non-binary variables).}
		\end{tabular}
	\end{table}

\subsection{Additional results}\label{subsec:app_addres}

	\begin{table}[H]
		\centering
		\caption{Effect of ability ratio on performance, sensitivity towards past and future considerations} 
		\label{apptab:table_main_perf_sens}
		\begin{tabular}{l r r r r r r r r } 
			\toprule \toprule
			\textbf{}                    & ~~ & (1)    & ~~ &  (2)   & ~~ &  (3)  & ~~ &  (4)       \\ 
			\midrule
            \multicolumn{9}{l}{\textit{Panel A: Underdogs's performance}} \vspace{0.7em}\\ 
		Current ability ratio$^1$ ~~~~~~   & &  -14.938***   & & -14.672*** & & -14.722*** & & -14.687***      \\
			                               & &   (3.005)     & & (3.229)    & &  (3.041)   & &  (3.244)        \\
			Prize money at stake$^4$       & &   0.378***    & &  0.616*    & &   0.283**  & &  0.511          \\
			(current round)                & &   (0.115)     & & (0.334)    & &  (0.128)   & &  (0.373)        \\
			Prize money at stake$^4$       & &               & &  -0.192    & &            & &  -0.204         \\
			(future rounds)                & &               & &  (0.262)   & &            & &  (0.262)        \\
			Expected ability$^3$           & &               & &  0.029     & &            & &   0.027         \\
			(next opponent)                & &               & &  (0.044)   & &            & &  (0.044)        \\
			Favorite legs played$^2$       & &               & &            & &  0.013     & &  0.004          \\
			                               & &               & &            & &   (0.023)  & &  (0.024)        \\
			Underdog legs played$^2$       & &               & &            & & 0.016      & &  0.014          \\
			                               & &               & &            & &  (0.023)   & &  (0.026)        \\
			\midrule
            \multicolumn{9}{l}{\textit{Panel B: Favorite's performance}} \vspace{0.7em} \\ 
			Current ability ratio$^1$      & &  5.737**      & & 6.651**    & &  4.983**   & &  5.986**        \\
			                               & &  (2.571)      & & (2.846)    & &  (2.527)   & &  (2.776)        \\
			Prize money at stake$^4$       & &  0.241*       & &  0.148     & &    0.211   & &  0.060          \\
			(current round)                & &  (0.127)      & &  (0.445)   & &  (0.140)   & &  (0.516)        \\
			Prize money at stake$^4$       & &               & &  -0.217    & &            & &  -0.207         \\
			(future rounds)                & &               & &   (0.310)  & &            & &  (0.317)        \\
			Expected ability$^3$           & &               & &  -0.060*   & &            & &   -0.063*       \\
			(next opponent)                & &               & &  (0.033)   & &            & &  (0.032)        \\
			Favorite legs played$^2$       & &               & &            & &  -0.029    & &  -0.017         \\
			                               & &               & &            & &   (0.024)  & &  (0.026)        \\
			Underdog legs played$^2$       & &               & &            & & 0.037      & &  0.032          \\
			                               & &               & &            & &  (0.025)   & &  (0.028)        \\
            \midrule
			Stage FE                       & &  x         & &  x         & &  x         & &  x      \\ 
            Tournament-by-Year FE          & &  x         & &  x         & &  x         & &  x      \\
			Individual FE                  & &  x         & &  x         & &  x         & &  x      \\
			All Covariates                 & &  x         & &  x         & &  x         & &  x      \\
			\midrule 
            N                              & &   4776     & &   4448     & &   4776     & &   4448  \\ 
			\bottomrule \bottomrule
			\multicolumn{9}{l}{\footnotesize Notes: *, **, and *** represents statistical significance at the 10 \%, 5 \%, and 1 \%, respectively. \textit{All Covariates}}\\
			\multicolumn{9}{l}{\footnotesize ~~~~~~~~~    include information on who starts the contest, if the favorite performs at home, and if the underdog}\\
			\multicolumn{9}{l}{\footnotesize ~~~~~~~~~     performs at home. FE = fixed effects. $^{1}$ Ratio is the favorite / underdog ability (measured as 3-darts }\\
			\multicolumn{9}{l}{\footnotesize ~~~~~~~~~    average over the past 2 years). $^{2}$ Number of legs played in the tournament before this contest.}\\
			\multicolumn{9}{l}{\footnotesize ~~~~~~~~~    $^{3}$ If the next opponent is known, this is their ability. Otherwise, it is the ability of the favorite of the  }\\
			\multicolumn{9}{l}{\footnotesize ~~~~~~~~~    contest that will determine the next opponent. $^{4}$ Prize money is measured as the proportion of the total  }\\
   			\multicolumn{9}{l}{\footnotesize ~~~~~~~~~   tournament purse the contestants can win in this round / future rounds. Linear regression. Standard }\\
			\multicolumn{9}{l}{\footnotesize ~~~~~~~~~  errors are clustered at the individual level.}
		\end{tabular}
	\end{table}

\noindent

\begin{figure}[H]
		\centering
		\caption{Favorite win - subsample analysis}
		\label{fig:het_favw}
   			\includegraphics[width=0.7\textwidth]{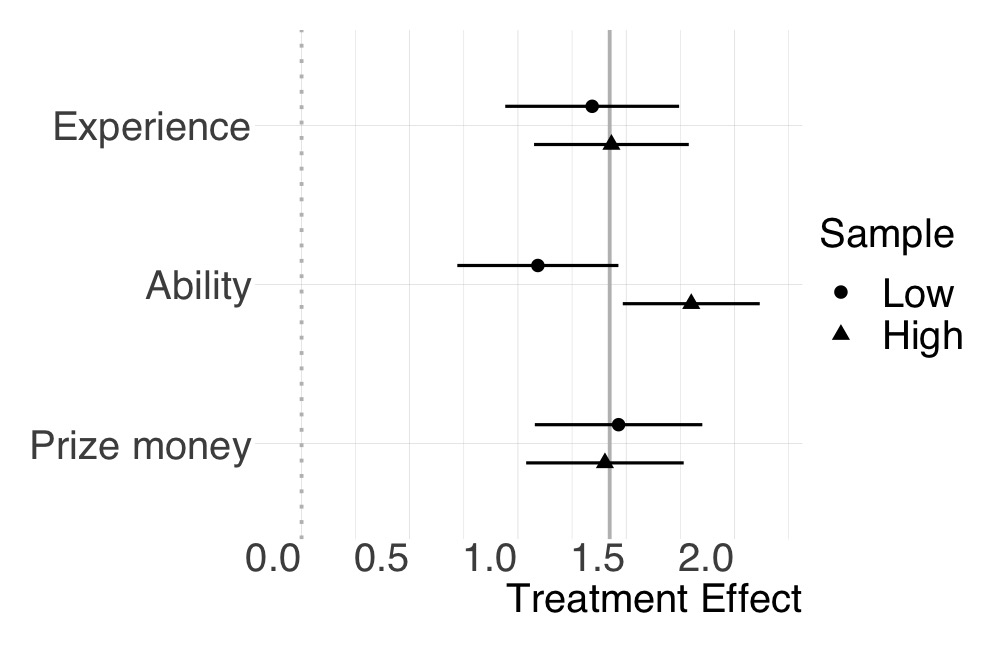}\\ 
		\footnotesize Notes: Linear regression on subsamples for low, i.e., below median, (circle) and high, i.e., above median,  \\
	\footnotesize ~~~~~~~~~~ (triangle) values of the Favorites' three heterogeneity variables. Specifications based on Table \ref{tab:table_main_perf},     ~~ \\
	\footnotesize ~~~~~~~~~~~ Column 3. The grey vertical lines indicate the average treatment effect. Whiskers mark the 90 \%   ~~\\
 	\footnotesize ~~~~~~~  confidence intervals.  \textit{Experience} is measured by years of playing darts. \textit{Ability} by the world  ~~~~~ \\
   	\footnotesize ~~~~~~~~~ ranking. \textit{Prize money} is the total tournament purse, standardized to account for increasing prize   \\ 
    \footnotesize ~ money over the years. For more details, see Section \ref{sec:data}). ~~~~~~~~~~~~~~~~~~~~~~~~~~~~~~~~~~~~~~~~~~~~~~~ \\
	\end{figure}

\newpage

    \begin{figure}[H]
		\centering
		\caption{Favorite win, shadow effect}
		\label{fig:app_fwin_future}
   			\includegraphics[width=0.7\textwidth]{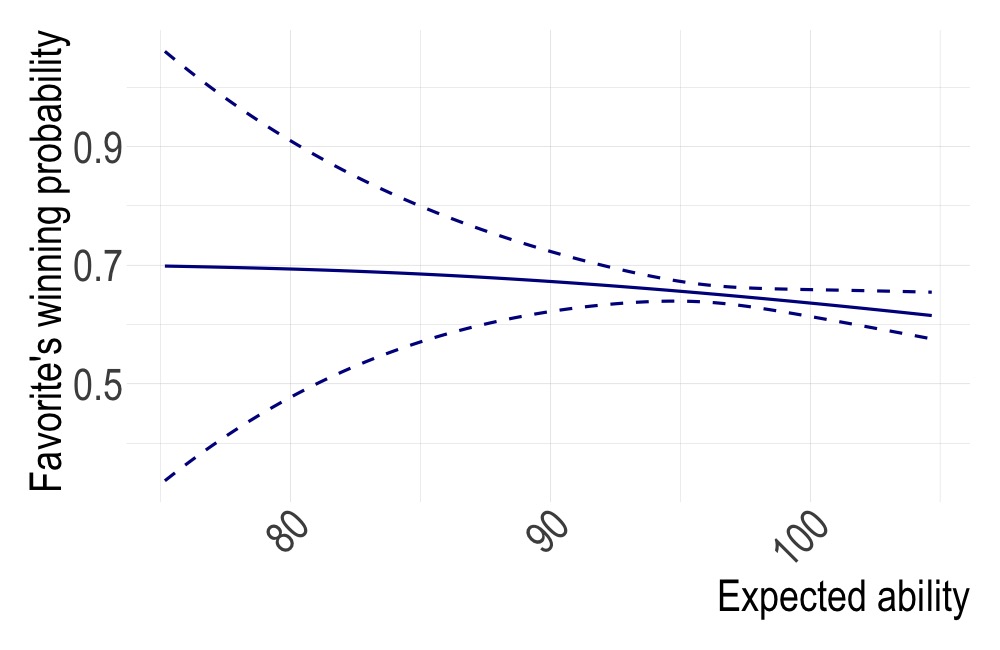}\\ 
		\footnotesize Notes: Non-parametric kernel regression. Based on Table \ref{tab:table_app_fut}, Column 2.  \\
		\footnotesize ~~~~~~~~~~ The blue line shows the expected winning probability for different \\
  		\footnotesize ~~~~~~~~   levels of the expected ability of the next opponent. Dashed lines  \\
    	\footnotesize ~   show the 90\% confidence intervals.  ~~~~~~~~~~~~~~~~~~~~~~~~~~~~~~~
	\end{figure}

\newpage

	\begin{table}[H]
		\centering
		\caption{Shadow effect -- 2SLS first stage} 
		\label{tab:table_app_2sls}
		\begin{tabular}{l r } 
			\toprule
            \toprule
			Opponent known before      & -2.031       \\
			                           & (0.108)      \\
			                           & [-18.75]     \\
			\midrule
			Stage FE    &     x        \\ 
			Tournament-by-Year FE      &     x        \\ 
			\midrule 
            N                          &   4448       \\ 
			\bottomrule
            \bottomrule
			\multicolumn{2}{l}{\footnotesize Notes: First stage estimation for Table \ref{tab:table_main_fut}. The outcome }\\
            \multicolumn{2}{l}{\footnotesize ~~~~~~~~~ variable is the expected ability of the opponent in }\\
            \multicolumn{2}{l}{\footnotesize ~~~~~~~~~  the next stage. Standard error (clustered on the  }\\
            \multicolumn{2}{l}{\footnotesize ~~~~~~~~~  individual level) in parenthesis. T-value in brackets.}\\
            \multicolumn{2}{l}{\footnotesize ~~~~~~~~~  FE = fixed effects.}
		\end{tabular}
	\end{table}

\newpage

\begin{table}[H]
		\centering
		\caption{Effect of ability ratio on performance of the higher-ability contestant by strength of future opponents} 
		\label{apptab:table_resource}
		\begin{tabular}{l r r r r r } 
			\toprule
			\toprule
			& \multicolumn{1}{c}{\textbf{Favorite Performance}} & ~~~ &  \multicolumn{3}{c}{\textbf{ Favorite Performance in }} \\
   			& \multicolumn{1}{c}{\textbf{within contest}} & ~~~ &  \multicolumn{1}{c}{\textbf{~~first half~~}} & ~ & \multicolumn{1}{c}{\textbf{second half}} \\
            \cline{2-2} \cline{4-6} \rule{0pt}{3ex}   
			        & (1)        & ~ & (2)   & ~ & (3)     \\ 
			\midrule 		
			\multicolumn{6}{l}{\textit{Low ability next opponent}}      \\
            \rule{0pt}{0.01ex}\\
			Ability ratio            & 8.628**   & & 11.633***   & & 4.153   \\
			                         & (3.596)   & & (4.181)     & & (5.401)  \\         
            \rule{0pt}{0.3ex}\\
            \multicolumn{6}{l}{\textit{High ability next opponent}}      \\
            \rule{0pt}{0.01ex}\\
			Ability ratio            & 2.775   & & 14.180**   & & -9.613**   \\
			                         & (3.884)   & & (6.023)     & & (4.389)  \\
            \rule{0pt}{0.3ex}\\
			\midrule
			Stage FE ~~ & x  & & x & & x  \\ 
			Tournament-by-Year FE    & x  & & x & & x  \\ 
			Individual FE            & x  & & x & & x  \\ 
			All Covariates           & x  & & x & & x  \\
			\midrule 
			N                        &  4776    & &  4776   & &   4776   \\ 
			\bottomrule
			\bottomrule
			\multicolumn{6}{l}{\footnotesize Notes: *, **, and *** represents statistical significance at the 10 \%, 5 \%, and 1 \%, respectively. Linear Regression.}\\
			\multicolumn{6}{l}{\footnotesize ~~~~~~~~~ Standard errors are clustered at the individual level. Low-ability next opponents are defined as below }\\
			\multicolumn{6}{l}{\footnotesize ~~~~~~~~~ median, high ability next opponents are defined as above median ability of the next opponent.}
		\end{tabular}
	\end{table}

 \newpage

\subsection{Robustness}\label{subsec:app_robust}

In Section \ref{subsec:cont_het}, we showed that the effect of increased ability heterogeneity on performance differs when focusing on higher-ability or lower-ability contestants. This section aims to confirm the consistency of this result across various specifications.\par

A potential concern regarding our results is the definition of the treatment variable. Section \ref{subsec:cont_het} defines the treatment as the ability ratio between the higher-ability contestant and the lower-ability contestant based on their past performances. There are two potential concerns: 1) Dependence on the functional form of the treatment variable and 2) Measurement error in the treatment variable. To address these concerns, we conduct three robustness checks. First, in Column 1 of Table \ref{apptab:table_rob_new}, we demonstrate that our results remain unchanged when we measure heterogeneity using a different functional form, i.e., the difference in the logarithm of the past performances between contestants instead of the ability ratio.\par

Second, our treatment variable could have a potential measurement error, given that the contestant's ability is approximated by the performance over the past two years.
Thus, we redefine our treatment variable based on bookmakers' odds, which are determined just before the contest starts.\footnote{However, this approach has a caveat. The bookmakers' odds are recorded after the starting contestants are known, meaning that the advantage of starting the contest is already incorporated into the odds \cite{Goller.2023}. Consequently, there is some misspecification in identifying the ex-ante higher-ability contestant and the ex-ante ability ratio. Additionally, since the bookmakers' prediction model is unknown, it is possible that the ability ratio itself was used to formulate the prediction. Therefore, we do not use the bookmaker odds in our preferred specification.} The results presented in Column 2 of Table \ref{apptab:table_rob_new} align with our main specifications. Third, we present a robustness check in which we remove the highest 1\% of the treatment values, which helps to address potential outliers in the data. We present the results in Column 3 of Table \ref{apptab:table_rob_new}. Removing these observations increases the point estimates slightly for the higher-ability contestants.\par 

Our identification strategy relies on the tournament draw, which generates "quasi-random" variation in opponent ability given a contestant's own ability. However, a potential concern arises regarding the adequacy of controlling for individual ability using individual fixed effects, as contestant ability might vary over time. To address this concern, we implement several robustness checks. First, we incorporate individual-by-year fixed effects as a control variable instead of individual fixed effects. By doing so, we account for potential changes in individual ability over time. The robustness check in Column 5 of Table \ref{apptab:table_rob_new} demonstrates that our main findings remain robust even when employing individual-by-year fixed effects. Furthermore, we conduct an additional robustness check, in which we directly control for ability by including information on past ability over the last two years as a covariate (Table \ref{apptab:table_rob_new}, Column 6). \par

Another concern about our identification strategy is that while the assignment of contestants via the tournament draw generates "quasi-random" variation in the opponents' ability and thus contestant heterogeneity in the first stage of the tournament, the heterogeneity among contestants in the upcoming stages is to some degree, but potentially not entirely exogenous. To address this concern, we show in Column 4 of Table \ref{apptab:table_rob_new} that our results are unchanged when we drop the last two stages of each tournament in our sample.\par

A final concern may relate to our outcome variable. Our main analysis uses the 3-darts average over the first nine darts as our performance measure. We chose this measure for the following reasons: 1) We can observe the performance in the same task for the higher-ability and lower-ability contestants 2) The dominant strategy within the first nine darts is to score the maximum number of points possible and there are few strategic considerations. Strategic choices solely arise as soon as a contestant ’sets’ the remaining number of points to be scored to a certain value by a throw that is not aimed at the maximum point yield. However, even though only a very small number of legs end with nine darts, there exists the possibility of finishing a leg with exactly nine darts or that the ninth dart is already strategically played to ease the "double out". Therefore, we conduct a robustness check in which we redefine our outcome variable. In Column 7 of Table \ref{apptab:table_rob_new}, we show results using the 3-darts average over the first six darts (two moves) as an alternative outcome variable. Our results are robust to this alternative definition of the outcome variable.\par

Lastly, we analyze alternative outcome variables in Table \ref{apptab:table_base_1_180}. These variables include the number of times an individual scores more than 100, 140, and precisely 180 points with three darts per leg. Professional darts commonly use these measures to evaluate contestants' performance, with 180 points being the highest attainable score. By examining these alternative metrics, we verify the consistency of our main findings across different performance indicators. Table \ref{apptab:table_base_1_180} confirms our main findings. The lower-ability contestant experiences substantial adverse effects across all outcome variables. Conversely, the higher-ability contestant has positive effects in all specifications, with statistical significance observed only in the case of scoring more than 100 points per move.\par

We conclude that our main results are robust to different outcomes, treatment definitions, and alternative ways of controlling the individuals' ability. \par

\begin{landscape}

	\begin{table}[H]
		\centering
		\caption{Robustness checks} 
		\label{apptab:table_rob_new}
		\begin{tabular}{l r r r r r r r r r r} 
			\toprule \toprule
            & \multicolumn{2}{c}{Treatment} & & \multicolumn{2}{c}{Sample} & & \multicolumn{2}{c}{Controls} & & Outcome \\
            \cline{2-3} \cline{5-6} \cline{8-9} \cline{11-11}
    \rule{0pt}{3ex}   
			& \multicolumn{1}{c}{(1)}        & \multicolumn{1}{c}{(2)} & & \multicolumn{1}{c}{(3)}   & \multicolumn{1}{c}{(4)} & & \multicolumn{1}{c}{(5)} & \multicolumn{1}{c}{(6)}  & & \multicolumn{1}{c}{(7)} \\
                      \rule{0pt}{3ex}
    		& \multicolumn{1}{c}{log spec.$^1$} & \multicolumn{1}{c}{odds spec.} & & \multicolumn{1}{c}{w/o outliers}   & \multicolumn{1}{c}{w/o (s.)finals} & & \multicolumn{1}{c}{ind x yr FE} & \multicolumn{1}{c}{ability$^2$}  & & \multicolumn{1}{c}{6 darts av.} \\
			\midrule
			\multicolumn{11}{c}{\textit{Panel A: Performance in contest}}      \\
			\multicolumn{11}{l}{\textit{A.1 Favorite's performance}}      \\
            \rule{0pt}{0.01ex}\\
			Ability ratio ~~~~ & 6.392**   & 0.297*** & & 7.145**  & 5.571**  & & 5.584*  & 3.934    & & 7.663***   \\
			                         & (2.843)   & (0.045)  & & (2.859)  & (2.739)  & & (3.291) & (2.464)  & & (2.919)  \\
            \rule{0pt}{0.3ex}\\
			\multicolumn{11}{l}{\textit{A.2 Underdog's performance}}  \\
            \rule{0pt}{0.01ex}\\
			Ability ratio            & -16.351*** & -0.209*** & & -13.662*** & -15.141*** & & -11.156** & -9.725**  & & -15.263***  \\
			                         & (3.279)    & (0.059)  & & (3.211)    & (3.058)   & & (5.397)   & (3.737)     & & (3.574)  \\
            \rule{0pt}{0.3ex}\\
			\midrule
   			\multicolumn{11}{c}{\textit{Panel B: Performance in first half of contest}}      \\
			\multicolumn{11}{l}{\textit{B.1 Favorite's performance}}      \\
            \rule{0pt}{0.01ex}\\
			Ability ratio            & 13.686***   & 0.377*** & & 14.908***  & 11.877***  & & 14.412***  & 10.967***    & & 15.103***  \\
			                         & (3.869)     & (0.077)  & & (3.775)    & (3.618)   & & (4.180)    & (3.508)       & & (4.339) \\
            \rule{0pt}{0.3ex}\\
			\multicolumn{11}{l}{\textit{B.2 Underdog's performance}}  \\
            \rule{0pt}{0.01ex}\\
			Ability ratio            & -14.674*** & -0.277*** & & -13.230*** & -13.257*** & & -13.441** & -10.751**  & & -10.931***  \\
			                         & (4.674)    & (0.078)  & & (4.514)    & (4.184)    & & (6.659)   & (4.881)     & & (4.704)  \\
            \rule{0pt}{0.3ex}\\
			\midrule
			N                        &  4776    & 4769 & &  4728   & 4433 & &  4776 &  4776    & &   4776   \\ 
			\bottomrule \bottomrule
			\multicolumn{11}{l}{\footnotesize Notes: *, **, and *** represents statistical significance at the 10 \%, 5 \%, and 1 \%, respectively. Linear Regression. All regressions include stage,}\\
			\multicolumn{11}{l}{\footnotesize ~~~~~~~~~   tournament, and individual fixed effects (FE), and all covariates (as in Table \ref{tab:table_main_perf}, Column 3); unless otherwise noted. Standard errors are }\\
            \multicolumn{11}{l}{\footnotesize ~~~~~~~~~ clustered at individual level. $^1$ The treatment is specified as differences instead of ratios. $^2$ Controlled for ability instead of Individual FE.}
		\end{tabular}
	\end{table}

\end{landscape}

	\begin{table}[H]
		\centering
		\caption{Results for within contest performance, 3-darts scores per leg} 
		\label{apptab:table_base_1_180}
		\begin{tabular}{l r r r r r } 
			\toprule
			\toprule
			& \multicolumn{1}{c}{\textbf{$>$100 per leg}} & ~~~ &  \multicolumn{1}{c}{\textbf{$>$140 per leg}} & ~~~ &  \multicolumn{1}{c}{\textbf{180 per leg}} \\
            \cline{2-2} \cline{4-4} \cline{6-6} \rule{0pt}{3ex}   
			        & (1)        & ~ & (2)   & ~ & (3)     \\ 
			\midrule
			\multicolumn{6}{l}{\textit{Panel A: Favorite's performance}}      \\
            \rule{0pt}{0.01ex}\\
			Ability ratio            & 0.333***    & & 0.111     & & 0.045   \\
			                         & (0.111)     & & (0.077)   & & (0.056)  \\
            \rule{0pt}{0.3ex}\\
			\multicolumn{6}{l}{\textit{Panel B: Underdog's performance}}  \\
            \rule{0pt}{0.01ex}\\
			Ability ratio            & -0.578***   & & -0.394*** & & -0.162***  \\
			                         & (0.115)     & & (0.099)   & & (0.053)     \\
            \rule{0pt}{0.3ex}\\
			\multicolumn{6}{l}{\textit{Panel C: Mean performance}}   \\
            \rule{0pt}{0.01ex}\\
			Ability ratio            & -0.172      & & -0.142*   & & -0.094*  \\
			                         & (0.109)     & & (0.083)   & & (0.050)     \\
            \rule{0pt}{0.3ex}\\
			\midrule
			Stage FE ~~ & x  & & x & & x  \\ 
			Tournament-by-Year FE    & x  & & x & & x  \\ 
			Individual FE            & x  & & x & & x  \\ 
			All Covariates           & x  & & x & & x  \\
			\midrule 
			N                        &  4776    & &  4776   & &   4776   \\ 
			\bottomrule
			\bottomrule
			\multicolumn{6}{l}{\footnotesize Notes: *, **, and *** represents statistical significance at the 10 \%, 5 \%, and 1 \%, respectively. }\\
			\multicolumn{6}{l}{\footnotesize ~~~~~~~~~ Linear Regression. Standard errors are clustered at the individual level. The outcome variables }\\
   			\multicolumn{6}{l}{\footnotesize ~~~~~~~~~ count the number of times an individual scores more than 100 (col. 1), more than 140 (col. 2)}\\
   			\multicolumn{6}{l}{\footnotesize ~~~~~~~~~ and exactly 180 (col. 3) with three darts in a leg. FE = fixed effects.}
		\end{tabular}
	\end{table}

\subsection{Full result}\label{subsec:app_full_tabs}

\subsubsection{Results with regard to current opponents}

	\begin{table}[H]
		\centering
		\caption{First and second half performance -- full table} 
		\label{tab:table_app_half}
		\begin{tabular}{l r r r r r} 
			\toprule \toprule
			& \multicolumn{2}{c}{\textbf{Favorite}} & &  \multicolumn{2}{c}{\textbf{Underdog}} \\
			& \multicolumn{2}{c}{\textbf{performance in }} & &  \multicolumn{2}{c}{\textbf{performance in }} \\
   			& \multicolumn{1}{c}{\textbf{~~first half~~}} & \multicolumn{1}{c}{\textbf{second half}} &  &  \multicolumn{1}{c}{\textbf{~~first half~~}} & \multicolumn{1}{c}{\textbf{second half}} \\
            \cline{2-3} \cline{5-6} \vspace{0.7em}\\ 
			Ability ratio$^1$          & 12.303*** & -2.035     & & -13.349*** &  -17.198*** \\
			                           & (3.492)   &  (3.541)   & &  (4.088)   &   (3.939)   \\
			Favorite starts first leg  &   0.152   &   0.297    & &  -0.436    &  -0.921***  \\
			                           &  (0.312)  &  (0.367)   & &  (0.285)   &   (0.322)   \\
			Favorite world ranking     & -5.086    &   3.171    & & -3.056*    &   1.205     \\
			                           &  (3.677)  &  (4.116)   & &  (1.763)   &   (2.137)   \\
			Underdog world ranking     &  -0.747   &   0.231    & & -3.122*    &   -3.350    \\
			                           &  (1.296)  &  (1.394)   & &  (1.781)   &   (2.096)   \\
			Favorite experience$^{2}$  &  2.379**  & 2.310*     & & 0.026      &   0.027     \\
			                           &   (0.970) &  (1.183)   & &  (0.017)   &   (0.019)   \\
			Underdog experience        &   0.019   &   -0.004   & &  3.676***  &   0.717    \\
			                           &  (0.013)  &  (0.017)   & &  (1.323)   &   (1.474)   \\
			Favorite at home event     &  -0.249   &  1.177     & &  0.999     &   1.106     \\
			                           &  (0.712)  &  (0.933)   & &  (0.750)   &   (0.872)   \\
			Underdog at home event     &  -0.210   &  -0.310    & & -0.129     &    0.760    \\
			                           &   (0.794) &  (0.780)   & &  (0.940)   &   (1.043)   \\
			\midrule
			Stage FE    &     x     &  x         & &  x         & x           \\ 
			Tournament-by-Year FE      &     x     &  x         & &  x         & x           \\ 
			Individual FE              &    x      &  x         & &  x         &  x         \\

			\midrule 
            N                          &   4776    &   4776     & &   4776     &    4776     \\ 
			\bottomrule \bottomrule
			\multicolumn{6}{l}{\footnotesize Notes: *, **, and *** represents statistical significance at the 10 \%, 5 \%, and 1 \%, respectively. }\\
			\multicolumn{6}{l}{\footnotesize ~~~~~~~~~ $^{1}$ Ratio is favorite / underdog, ability measured as 3-darts average over the past two years. }\\
			\multicolumn{6}{l}{\footnotesize ~~~~~~~~~ $^{2}$ Experience is measured as number of years playing darts. FE = fixed effects.}\\
			\multicolumn{6}{l}{\footnotesize ~~~~~~~~~ }
		\end{tabular}
	\end{table}

\subsubsection{Contest outcomes}

	\begin{table}[H]
		\centering
		\caption{Contest outcomes -- full table} 
		\label{tab:table_app_contest}
		\begin{tabular}{l r r r r r r} 
			\toprule \toprule
			& \multicolumn{3}{c}{\textbf{~~~~Favorite~~~~}} & & \multicolumn{2}{c}{\textbf{~~~~~~Legs~~~~~~}} \\
			& \multicolumn{3}{c}{\textbf{~~~~win~~~~}} & & \multicolumn{2}{c}{\textbf{needed}} \\
            \cline{2-4} \cline{6-7}  \vspace{0.01em}\\ 
			\textbf{}                   & (1)      &  (2)    &  (3)   & &  (4)   &  (5)    \\ 
			\midrule
			Ability ratio$^1$          &   1.420***  & 1.423***  & 1.445***  & & -0.346*** &   -0.346*** \\
			                           &   (0.156)   & (0.159)   & (0.143)   & & (0.053)   &  (0.053)    \\
			Favorite world ranking     &    0.155    & 0.152     & -0.097    & & -0.003    &   -0.002    \\
			                           &   (0.110)   & (0.113)   & (0.079)   & & (0.052)   &  (0.052)    \\
			Underdog world ranking     &    0.174*** & 0.171***  & 0.160***  & & 0.016     &   0.015     \\
			                           &   (0.049)   & (0.049)   & (0.045)   & & (0.017)   &  (0.016)    \\
			Favorite starts first leg  &             & 0.075***  & 0.081***  & &           & -0.009**    \\
			                           &             & (0.017)   & (0.014)   & &           &  (0.005)    \\
			Favorite experience$^{2}$  &             & -0.042    & -0.002**  & &           &  0.004      \\
			                           &             & (0.102)   & (0.001)   & &           &  (0.025)    \\
			Underdog experience        &             & 0.001     & 0.001     & &           &  -0.000     \\
			                           &             & (0.001)   & (0.001)   & &           &  (0.000)    \\
			Favorite at home event     &             & 0.024     & 0.035     & &           &  0.016*     \\
			                           &             & (0.034)   & (0.035)   & &           &  (0.009)    \\
			Underdog at home event     &             & 0.017     & 0.013     & &           &  -0.007     \\
			                           &             & (0.031)   & (0.029)   & &           &  (0.011)    \\
			Favorite ability           &             &           & 0.012***  & &           &             \\
			                           &             &           & (0.002)   & &           &             \\
			\midrule
			Stage FE    &   x         & x         &           & & x         & x           \\ 
			Tournament-by-Year FE      &   x         & x         &           & & x         & x           \\ 
			Individual FE              &   x         & x         &           & & x         & x           \\
			\midrule 
            N                          &   4776      & 4776      & 4776      & & 4776      &   4776      \\ 
			\bottomrule \bottomrule
			\multicolumn{7}{l}{\footnotesize Notes: *, **, and *** represents statistical significance at the 10 \%, 5 \%, and 1 \%, respectively. }\\
			\multicolumn{7}{l}{\footnotesize ~~~~~~~~~ $^{1}$ Ratio is favorite / underdog, ability measured as 3-darts average over the past 2 years. }\\
			\multicolumn{7}{l}{\footnotesize ~~~~~~~~~ $^{2}$ Experience is measured as number of years playing darts. FE = fixed effects.}\\
			\multicolumn{7}{l}{\footnotesize ~~~~~~~~~ In Column 2, both underdog and favorite fixed effects are used.}
		\end{tabular}
	\end{table}

\subsubsection{Full result table for head start}

	\begin{table}[H]
		\centering
		\caption{Underdog head start -- full table} 
		\label{tab:table_app_head}
		\begin{tabular}{l r r r r r r r} 
			\toprule
            \toprule
			& \multicolumn{1}{c}{\textbf{~~~Favorite~~~}} & ~~ &  \multicolumn{5}{c}{\textbf{Performance}}  \\
            \cline{4-8}
   			& \multicolumn{1}{c}{\textbf{win}} &  &  \multicolumn{1}{c}{\textbf{Favorite}} & & \multicolumn{1}{c}{\textbf{Underdog}} & & \multicolumn{1}{c}{\textbf{Average}}\\
            \cline{2-2} \cline{4-4} \cline{6-6} \cline{8-8} \vspace{0.7em}\\  
            
			Underdog head start  ~~~  & -0.075***    & & -0.228    & & 0.688***  & & 0.277    \\
			                         & (0.017)      & & (0.280)   & & (0.202)   & & (0.169)  \\
			Ability ratio            & 1.423***     & &  5.738**  & & -15.075*** & & -4.506  \\
			                         & (0.159)      & & (2.569)   & & (3.006)   & & (2.790)  \\
			Favorite rank            & 0.152        & & -0.920    & & -1.078    & & -1.593   \\
			                         & (0.113)      & & (1.672)   & & (1.257)   & & (1.476)  \\
			Underdog rank            &  0.171***    & & -0.238    & & -3.297**  & & -1.236   \\
			                         & (0.049)      & & (1.000)   & & (1.547)   & & (1.042)  \\
			Favorite experience      & -0.042       & & 2.373***  & & 0.028**   & & 1.025    \\
			                         & (0.102)      & & (0.689)   & & (0.013)   & & (0.839)  \\
			Underdog experience      &  0.001       & &  0.008    & & 2.257**   & & 0.977*** \\
			                         & (0.001)      & & (0.012)   & & (1.049)   & & (0.320)  \\
			Favorite home            &  0.024       & &  0.417    & & 1.062**   & & 0.838*   \\
			                         & (0.034)      & & (0.601)   & & (0.501)   & & (0.443)  \\
			Underdog home            &  0.017       & & -0.265    & & 0.338     & & 0.006    \\
			                         & (0.031)      & & (0.626)   & & (0.814)   & & (0.508)  \\
            \rule{0pt}{0.3ex}\\
            \midrule
			Stage FE  & x  & & x & & x & & x  \\ 
			Tournament-by-Year FE    & x  & & x & & x & & x  \\ 
			Individual FE            & x  & & x & & x & & x  \\ 			 
			\bottomrule
            \bottomrule
			\multicolumn{8}{l}{\footnotesize Notes: *, **, and *** represents statistical significance at the 10 \%, 5 \%, and 1 \%, respectively. }\\
			\multicolumn{8}{l}{\footnotesize ~~~~~~~~~ Standard errors are clustered at the individual level or robust for average performance.}\\
			\multicolumn{8}{l}{\footnotesize ~~~~~~~~~ favorite and underdog FE for average performance. FE = fixed effects. Linear regression.}
		\end{tabular}
	\end{table}

\subsubsection{Results with regard to future opponents}

	\begin{table}[H]
		\centering
		\caption{Future contest opponents -- full table} 
		\label{tab:table_app_fut}
		\begin{tabular}{l r r r r r r r r} 
			\toprule \toprule
			& \multicolumn{2}{c}{\textbf{~~~Favorite~~~}} &  &  \multicolumn{5}{c}{\textbf{Performance}}  \\
            \cline{5-9}
   			& \multicolumn{2}{c}{\textbf{win}} &  &  \multicolumn{1}{c}{\textbf{Favorite}} & & \multicolumn{1}{c}{\textbf{Underdog}} & & \multicolumn{1}{c}{\textbf{Average}}\\
            \cline{2-3} \cline{5-5} \cline{7-7} \cline{9-9} \vspace{0.1em}\\
			\textbf{}                  & (1)       &  (2)        & & (3)        & & (4)        & & (5)     \\ 
			\midrule
			Expected ability$^1$       & -0.004*   & -0.003*     & &  -0.063*   & &   0.066    & & -0.019   \\
			                           & (0.002)   & (0.002)     & &  (0.032)   & &  (0.043)   & & (0.028)  \\
			Favorite world ranking     &  0.111    & -0.208**    & &  -1.932    & & -1.186     & & -2.796** \\
			                           &  (0.117)  & (0.104)     & &  (1.764)   & &  (1.377)   & & (1.243)  \\
			Underdog world ranking     & 0.251***  & 0.219**     & &   0.142    & & -3.194**   & & -1.223   \\
			                           &  (0.052)  & (0.048)     & &  (1.065)   & &  (1.556)   & & (1.231)  \\
			Favorite starts first leg  & 0.078***  & 0.083***    & &   0.237    & &  -0.603*** & & -0.206   \\
			                           &  (0.017)  & (0.014)     & &  (0.288)   & &  (0.217)   & & (0.211)  \\
			Favorite experience$^{2}$  &  -0.021   & -0.002      & & 3.370***   & & 0.030**    & & 1.777*   \\
			                           &   (0.100) & (0.001)     & &  (0.816)   & &  (0.013)   & & (0.921)  \\
			Underdog experience        &   0.000   & 0.001       & &   0.011    & &  2.693**   & & 0.979*** \\
			                           &  (0.001)  & (0.001)     & &  (0.012)   & &  (1.154)   & & (0.336)  \\
			Favorite at home event     &   0.026   & 0.022       & &  0.296     & &  1.115**   & & 0.752    \\
			                           &  (0.037)  & (0.036)     & &  (0.602)   & &  (0.555)   & & (0.550)  \\
			Underdog at home event     &   0.029   & 0.018       & &   0.024    & &  0.235     & & 0.010    \\
			                           &   (0.035) & (0.033)     & &  (0.647)   & &  (0.846)   & & (0.673)  \\
			\midrule
			Stage FE    &     x     &             & &  x         & &  x         & & x       \\ 
			Tournament-by-Year FE      &     x     &             & &  x         & &  x         & & x       \\
			Individual FE              &   x       &             & &  x         & & x          & & x       \\
			\midrule 
            N                          &   4776    &    4776     & &   4776     & &   4776     & & 4776    \\ 
			\bottomrule \bottomrule
			\multicolumn{9}{l}{\footnotesize Notes: *, **, and *** represents statistical significance at the 10 \%, 5 \%, and 1 \%, respectively. }\\
			\multicolumn{9}{l}{\footnotesize ~~~~~~~~~ $^{1}$ If the next opponent is known, this is their ability, otherwise it is the ability of the }\\
			\multicolumn{9}{l}{\footnotesize ~~~~~~~~~  favorite of the contest that will determine the next opponent. $^{2}$ Experience is measured }\\
			\multicolumn{9}{l}{\footnotesize ~~~~~~~~~ as number of years playing darts. FE = fixed effects.}
		\end{tabular}
	\end{table}

\end{document}